%% file: paper.tex
\documentclass[paper,notoc]{JHEP3}
\usepackage{axodraw4j}
\usepackage{amsmath} 
\usepackage{graphicx}
\usepackage{epsfig,cite}
\usepackage{booktabs}
\usepackage{caption}
\usepackage{subcaption}

\def\beq{\begin{equation}}   
\def\eeq{\end{equation}}
\def\bea{\begin{eqnarray}}  
\def\eea{\end{eqnarray}} 
\def\nn{\nonumber}
\def\eps{\epsilon}
\def\n3lo{$\mathrm{N^3LO}$}

\newcommand{\ord}[0]{\mathcal{O}}

\newcommand{\DD}{\mathcal{D}}
\newcommand{\Li}{\mathrm{Li}}
\newcommand{\dd}{\mathrm{d}}

\allowdisplaybreaks[1]

\title{Scale dependence and collinear subtraction terms for Higgs production in gluon fusion at N3LO}

\author{Stephan Buehler\\
  Institute for Theoretical Physics, ETH Zurich,
  8093 Zurich, Switzerland\\
  E-mail: \email{buehler@itp.phys.ethz.ch}}

\author{Achilleas Lazopoulos\\
  Institute for Theoretical Physics, ETH Zurich,
  8093 Zurich, Switzerland\\
  E-mail: \email{lazopoli@phys.ethz.ch}}

\abstract{
The full, explicit, scale dependence of the inclusive \n3lo cross section for  single Higgs hadroproduction  is obtained by calculating the convolutions of collinear splitting kernels with lower-order partonic cross sections.
We provide results for all convolutions of splitting kernels and lower-order partonic cross sections to the order in $\epsilon$ needed for the full \n3lo
computation, as well as their expansions around the soft limit. We also discuss the size of the total scale uncertainty at \n3lo that can be anticipated with existing information.}


\preprint{}

\begin{document}

%%%%%%%%%%%%%%%%%%%%%%%%%%%%%%%%%%%%%%%%%%%%%%%%%%%%%%%%%%%%%%%%%%%%%%%%
% Shamelessly stolen from Thorsten's thohacks.sty
%%%%%%%%%%%%%%%%%%%%%%%%%%%%%%%%%%%%%%%%%%%%%%%%%%%%%%%%%%%%%%%%%%%%%%%%
\catcode`\@=11
\font\manfnt=manfnt
\def\Watchout{\@ifnextchar [{\W@tchout}{\W@tchout[1]}}
\def\W@tchout[#1]{{\manfnt\@tempcnta#1\relax%
  \@whilenum\@tempcnta>\z@\do{%
    \char"7F\hskip 0.3em\advance\@tempcnta\m@ne}}}
\let\foo\W@tchout
\def\dubious{\@ifnextchar[{\@dubious}{\@dubious[1]}}
\let\enddubious\endlist
\def\@dubious[#1]{%
  \setbox\@tempboxa\hbox{\@W@tchout#1}
  \@tempdima\wd\@tempboxa
  \list{}{\leftmargin\@tempdima}\item[\hbox to 0pt{\hss\@W@tchout#1}]}
\def\@W@tchout#1{\W@tchout[#1]}
\catcode`\@=12
%%%%%%%%%%%%%%%%%%%%%%%%%%%%%%%%%%%%%%%%%%%%%%%%%%%%%%%%%%%%%%%%%%%%%%%%

%\maketitle

%%%%%%%%%%%%%%%%%%%%%%%%%%%%%%%%%%%%%%%%%%%%%%%%%
%%%%% Introduction
%%%%%%%%%%%%%%%%%%%%%%%%%%%%%%%%%%%%%%%%%%%%%%%%%%
\section{Introduction}
\label{sec:introduction}

During the past year both multi-purpose experiments at CERN's Large hardon collider (LHC),
CMS~\cite{Chatrchyan:2012ufa} and Atlas~\cite{Aad:2012tfa}, have
observed a new boson with a mass of about 125 GeV, which is strongly believed to be the long-sought Higgs boson. The couplings
of the new boson to Standard Model (SM) particles are currently compatible to  the SM predictions with a minimal Higgs sector. Nevertheless, effects from physics beyond the Standard Model (BSM)
may reside in small deviations of the couplings from the SM values, effects that will be, to a certain extent, accessible with the increased statistics and energy reach of the LHC
high energy run starting in 2015.

The dominant production mode for the Higgs boson at the LHC is  gluon fusion,  accounting for about 90$\%$ of the
total production cross section at the observed mass of about 125 GeV. Indeed, the Higgs boson has, up to now, been observed in channels in which its production is gluon induced. Next-to-leading order (NLO) QCD corrections for gluon fusion, in the five-flavour heavy-quark effective theory (HQET)
were computed at the beginning of the 1990s~\cite{Dawson:1990zj,Djouadi:1991tka}. Since then NLO corrections in the full theory including top-mass, and top-bottom interference effects were calculated by~\cite{Graudenz:1992pv,Spira:1995rr,Anastasiou:2006hc,Aglietti:2006tp,Bonciani:2007ex}, and next-to-next-to-leading order (NNLO) corrections in HQET by~\cite{Harlander:2002wh,Anastasiou:2002yz,Ravindran:2003um}.
Electroweak corrections are also available at the NLO level~\cite{Aglietti:2004nj,Degrassi:2004mx,Actis:2008ug,Actis:2008ts}, and so are mixed QCD-EW corrections~\cite{Anastasiou:2008tj}, and EW corrections to higgs plus jet including top and bottom quark contributions~\cite{Brein:2010xj}. Recently the NNLO cross section for gluon induced higgs production in association to a jet was calculated, in a way that also allows for differential distributions to be produced~\cite{Boughezal:2013uia}. All available fixed order contributions to Higgs production via gluon fusion were recently included in the program {\tt iHixs}~\cite{Anastasiou:2011pi}, which, moreover, allows for the incorporation of BSM effects through modified Wilson coefficients within the effective theory approach. The latter has been explicitly shown to be an excellent approximation for Higgs masses below the top-antitop threshold~\cite{Marzani:2008az,Harlander:2009mq,Pak:2009dg}, and even more so for the Higgs boson at 125 GeV. Despite these advances and due to the slow perturbative convergence of the gluon fusion cross section, the remaining uncertainty due to variation in renormalisation and factorisation scale still
amounts to about $\sim 9\%$ for a $125$ GeV Higgs boson at the LHC with $8$ TeV centre-of-mass energy.  

Beyond fixed order, threshold resummation has been performed to NNLL accuracy by traditional resummation methods~\cite{Catani:2003zt} leading to a $\sim 7.5\%$ uncertainty~\cite{deFlorian:2012yg}, and within the SCET framework~\cite{Ahrens:2008qu,Ahrens:2008nc,Ahrens:2010rs} leading to a $\sim 4\%$ scale uncertainty. The latter is generally considered too optimistic.    

Information from the LHC high energy and high luminosity data set is projected to allow the determination of the Higgs couplings with precision of $\sim10\%$ or better~\cite{Peskin:2012we,Klute:2012pu,Klute:2013cx}. This uncertainty includes experimental systematics and statistics, but also errors from the determination of parton distribution functions and of the strong coupling, as well as theory systematics, the latter being the limiting factor in several cases. It is evident that a prerequisite to this goal is the reduction of the theory scale uncertainty to the $\sim 5\%$ level or lower. The question arises then, whether computing the cross section to the next order in perturbation theory, \n3lo, within the EFT approach, an admittedly formidable task, would achieve this goal. 

Information about certain \n3lo contributions has been available for several years. The three-loop, virtual contributions have been calculated and were part of the full \n3lo Higgs decay to gluons in~\cite{Baikov:2006ch}. However, disentangling the pure virtual contributions from this computation is not possible. 
The quark and gluon form-factors are known
up to three-loop order~\cite{Moch:2005tm,Baikov:2009bg,Lee:2010cga,Gehrmann:2010ue}. In~\cite{Moch:2005ky} the soft 'plus'-contributions to the \n3lo cross section were
derived using mass factorisation constraints. This allowed the authors of ~\cite{Moch:2005ky} to derive a soft approximation of the \n3lo cross section whose renormalisation scale dependence is rather mild, resulting in $~4\%$ renormalisation scale uncertainty (keeping the factorisation scale equal to the Higgs mass). Recently further attempts to modify the resummation procedure such that its prediction at fixed order better matches the threshold and high energy limits of the known fixed order results, were made~\cite{Ball:2013bra}, resulting in another soft approximant with a  scale uncertainty of $7\%$. It still remains true that without the full \n3lo expression, it is difficult to judge which of these prescriptions is closer to reality. 

Recently, some new ingredients of the full \n3lo cross section have appeared. In~\cite{Pak:2011hs,Anastasiou:2012kq}, the real-virtual and double-real master integrals
of the NNLO cross section have been calculated to higher orders in $\epsilon$. In \cite{Hoschele:2012xc}, the convolutions of collinear splitting kernels with
lower-order partonic cross sections have been computed, which is also an ingredient for our result and has been re-derived in this work. Very recently,
the soft limits of all master integrals appearing in triple-real radiation corrections (i.e. the emission of three additional partons) have been worked
out~\cite{Anastasiou:2013srw}.

In this paper, we compute the full dependence of the \n3lo cross section on the factorisation and renormalisation scales, which can be obtained from
lower-order results.  Furthermore, we provide the soft limits of all convolutions that we calculated, which may become useful when expanding the full
\n3lo corrections around threshold. In section \ref{sec:review} we review how the dependence on factorisation and renormalisation scales enters higher-order
calculations. In sections \ref{sec:ingredients} and \ref{sec:conv} we list the splitting kernels and partonic cross sections needed for our results and 
present the method used to compute their convolutions, respectively.
In section \ref{sec:results} we give results for the estimated scale uncertainty of the \n3lo gluon fusion cross section and conclude in section \ref{sec:Conclusions}.
%
%%%%%%%%%%%%%%%%%%%%%%%%%%%%%%%%%%%%%%%%%%%%%%%%%
%%%%% review
%%%%%%%%%%%%%%%%%%%%%%%%%%%%%%%%%%%%%%%%%%%%%%%%%%
\section{Sources of explicit scale dependence}
\label{sec:review}
Predictions for observable quantities in quantum field theory are independent of arbitrary scales, when calculated at all perturbative orders. The scale dependence of all predictions is an artefact of the truncation of the perturbative series, and is usually considered a measure of the effect of missing higher orders in any given computation. This dependence occurs explicitly, through terms in the final result that depend on logarithms involving the scale, and implicitly, through the running of $\alpha_s$ and the evolution of the parton distribution functions. In this section we describe the occurrence of the explicit scale dependence. 

Let us, for the moment, introduce only one scale, 
\begin{equation}
\mu_r=\mu_f\equiv\mu.
\end{equation} 
In dimensional regularisation the scale $\mu$ appears during renormalisation, when the bare coupling is replaced by the renormalised one,
\begin{equation}
\alpha_s^{(B)} \mapsto \alpha_s(\mu)\,\mu^{2\epsilon} \left(\frac{e^{\gamma_E}}{4\pi}\right)^\epsilon \, Z_\alpha \, ,
\end{equation} 
where we have chosen the $\overline{\text{MS}}$-scheme. $Z_\alpha$ is the renormalisation constant of the strong coupling,
\begin{equation}
Z_\alpha = 1 - a(\mu) \frac{\beta_0}{\epsilon} + a^2(\mu) \left(\frac{\beta_0^2}{\epsilon^2} - \frac{\beta_1}{2\epsilon} \right)
 + a^3(\mu) \left(-\frac{\beta_0^3}{\epsilon^3} + \frac{7\beta_0\beta_1}{6\epsilon} - \frac{\beta_2}{3\epsilon} \right) + \mathcal{O}(a^4)\, ,
\end{equation}
and the factor of $\mu^{2\epsilon}$ ensures that the coupling
and thus the action remain dimensionless in $D=4-2\epsilon$ dimensions as well. We define $a(\mu) \equiv \frac{\alpha_s(\mu)}{\pi}$ throughout the paper.

Divergences, of UV or IR nature, manifest themselves as poles in the regularisation parameter $\epsilon$. The leading divergences, $\epsilon^{-2n},\ldots,\epsilon^{-n-1}$ for the $n$-th order correction, vanish, among real and virtual contributions,  after renormalisation counter terms are included.  The remaining poles of the  UV-renormalized partonic cross section, starting from $\epsilon^{-n}$, vanish only after  subtraction of collinear counterterms.

Specifically, let us denote by $\hat{\sigma}_{ij}$ the partonic cross section after renormalisation\footnote{Note the $1/z$ factor in the definition of $\hat{\sigma}_{ij}$ that is necessary to make eq.~\ref{mother_convolution} work.} (which still contains divergences of infrared (IR) origin),
\beq
\hat{\sigma}_{ij} = \frac{1}{z} \left[ \frac{1}{2s} \int \left[ d\Phi \right]  \sum_{\text{spin},\text{col}}\left| \overline{M} \right|^2  + UV_{\text{counterterms}}\right] 
\label{sigma_hat_definition}
\eeq
The expansion of $\hat{\sigma}_{ij}$ can be written as 
\beq
\hat{\sigma}_{ij} = \sum_{n=0}^{\infty} a^n \hat{\sigma}_{ij}^{(n)} = \sum_{n=0}^{\infty} a^n \sum_{r=-n}^\infty \hat{\sigma}_{ij}^{(n,r)}\,\epsilon^r\,e^{L_f\epsilon}
\eeq
where we have written explicitly  the pole coefficients at every order in $a$ and the associated logarithms $L_f\equiv \log(\mu^2/s)$.  
The relation of  $\hat{\sigma}_{ij}$  to the total, inclusive cross section is given by convolution with the parton distribution functions, $f_i(x)$, and the collinear counter terms, $\Gamma_{ij}(x)$, by 

\begin{equation}
\sigma(\tau) =\tau \left(f_i\, \otimes \Gamma^{-1}_{ki} \, \otimes \, \hat{\sigma}_{kl} \, \otimes \, \Gamma^{-1}_{lj} \otimes \, f_j \right)(\tau) \, ,
\label{mother_convolution}
\end{equation}
where summation of repeated indices is implied and $\tau=m_h^2/S$ with $S$ the total centre-of-mass  energy of the collision. 

The convolution is defined by
\begin{equation}
\left(f\otimes g\right) (z) = \int_0^1 \mathrm{d}x \mathrm{d}y \, f(x)\,g(y)\,\delta(xy-z) \, ,
\end{equation}
and the collinear counter term reads
\begin{align}
\Gamma_{ij}(x)  = & \,\delta_{ij}\delta(1-x) - a(\mu) \frac{P^{(0)}_{ij}(x)}{\epsilon} \nonumber \\ 
 & + a^2(\mu) \Bigg\{ -\frac{1}{2\epsilon}P^{(1)}_{ij}(x) 
  + \frac{1}{2\epsilon^2} \left[ \left(P^{(0)}_{ik}\otimes P^{(0)}_{kj}\right)(x) + \beta_0 P^{(0)}_{ij}(x)\right] \Bigg\} \nonumber \\
 & + a^3(\mu)\Bigg\{ -\frac{1}{6\epsilon^3} \Big[ \left(P^{(0)}_{ik}\otimes P^{(0)}_{kl}\otimes P^{(0)}_{lj}\right)(x) + 3\beta_0\left(P^{(0)}_{ik}\otimes P^{(0)}_{kj}\right)(x) \nonumber \\
 & + 2\beta_0^2 P^{(0)}_{ij}(x) \Big] + \frac{1}{12\epsilon^2} \Big[3\left(P^{(0)}_{ik}\otimes P^{(1)}_{kj}\right)(x) + 3\left(P^{(1)}_{ik}\otimes P^{(0)}_{kj}\right)(x) \nonumber \\ 
 & + 4\beta_1 P^{(0)}_{ij}(x) + 4\beta_0P^{(1)}_{ij}(x)\Big] - \frac{1}{3\epsilon} P^{(2)}_{ij}(x) \Bigg\} +\ord(a^4) \, .
 \label{coll_subtraction_terms}
\end{align}
The $P^{(n)}_{ij}$ are the Altarelli-Parisi splitting kernels which govern the emission of collinear partons (see section \ref{sec:AP}).\\

Within the renormalized N$^n$LO partonic cross section, $\hat{\sigma}_{ij}$, the logarithmic dependence on the scale $\mu$ arises when residual poles of order up to $\epsilon^{-n}$ are multiplied with the expansion of the factor $\mu^{2\epsilon}/s^\epsilon$ 
(the $s^{-\epsilon}$ originating from the $d$-dimensional phase-space measure),
\begin{equation}
\frac{\mu^{2\epsilon}}{s^\epsilon} = 1 + \epsilon\log\left(\frac{\mu^2}{s}\right) + \frac{\epsilon^2}{2}\log^2\left(\frac{\mu^2}{s}\right) + \frac{\epsilon^3}{6}\log^3\left(\frac{\mu^2}{s}\right) + \ord(\epsilon^4) \, .
\end{equation}
These poles are required to cancel against the poles from the collinear counter terms convoluted with lower order partonic cross sections, see eq.~\eqref{mother_convolution} and \eqref{coll_subtraction_terms}. This requirement fixes the coefficients $\hat{\sigma}_{ij}^{(n,r)}$ for $-n-1<r < 0$ which are also the coefficients of the logarithmic terms. In summary, all contributions to the N$^n$LO cross section that are proportional to a power of $\log(\mu^2/s)$ can be obtained from calculating the convolutions of splitting kernels and lower-order, N$^{m<n}$LO partonic cross sections.

The computation of all combinations of splitting kernels and partonic cross sections relevant for the \n3lo corrections to the gluon fusion process is the main work of this publication.
We calculate also all pieces of higher orders in $\epsilon$ that will add to the finite part of the \n3lo corrections (but not necessarily to the scale dependent parts).

With the pole cancelation achieved, let us now define the finite, mass-renormalised partonic cross section
\begin{equation}
  \sigma_{ij}(\mu,z) =\left(  \Gamma^{-1}_{ki} \otimes \Gamma^{-1}_{lj} \otimes \hat{\sigma}_{kl} \right) (z) \, ,
\end{equation}
with $\hat{\sigma}_{km}$   now explicitly dependent on $\log(\mu^2 /s)$.
Alternatively, the relation above can be inverted,
\begin{equation}
  \hat{\sigma}_{ij} (z) = \left( \Gamma_{ki}\otimes\Gamma_{lj}\otimes\sigma_{kl}\right)(z) \, ,
\end{equation}
and solved for the highest order of $\sigma$ one is interested in. For example, at NLO, this yields (using $\Gamma^{(0)}_{ij}(z) = \delta_{ij}\delta(1-z)$)
\begin{equation}
  \sigma^{(1)}_{ij} = \hat{\sigma}^{(1)}_{ij} - \Gamma^{(1)}_{ki}\otimes\sigma^{(0)}_{kj} - \Gamma^{(1)}_{kj}\otimes\sigma^{(0)}_{ik} \, .
\end{equation}
This step-by-step procedure provides an additional test on the lower-order cross section, since a mistake in their mass-renormalisation
will result in uncancelled poles at a higher order. We provide results in this framework, i.e. our convolutions involve splitting kernels
and the finite partonic cross section $\sigma_{ij}$.

We can, then, set $\mu=\mu_f$ and use the renormalisation group equation for the strong coupling constant, 
\begin{equation}
\label{eq:rge}
\frac{\partial a(\mu)}{\partial \log(\mu^2)} = \beta(\mu)\,a(\mu) \quad \text{with} \quad \beta(\mu) = - \sum_{n=0} \beta_na^{n+1}_s(\mu) \, .
\end{equation}
 to change the scale at which $\alpha_s$ is evaluated in $\sigma_{ij}(\mu_r,\mu_f,z)$. 

The third-order expansion of $a(\mu_f)$ in terms of $a(\mu_r)$ reads
\begin{align}
\label{eq:ascaletrans}
a(\mu_f)  = & a(\mu_r) +  a^2(\mu_r) \beta_0 L 
               + a^3(\mu_r) \left[\beta_1 L +\beta_0^2 L^2 \right] + \nonumber \\
             & + a^4(\mu_r) \left[ \beta_2 L + \frac{5}{2}\beta_1\beta_0 L^2 + \beta_0^3 L^3 \right]  + \ord(a^5)\, ,
\end{align} 
with $L\equiv\log(\mu_r^2/\mu_f^2)$. 

The explicit logarithm $L$ essentially counters the running of the coupling constant up to the order considered, such that the 
effect of varying the unphysical scales is weakened for higher orders in perturbation theory, i.e. the perturbative series is converging to its all-order value, as can be 
seen by taking the total derivative of the partonic cross section,
\begin{equation}
\frac{\mathrm{d}\sigma_{ij}(\mu)}{\mathrm{d}\log(\mu^2)} = \frac{\partial\sigma_{ij}(\mu)}{\partial\log(\mu^2)} + \frac{\partial\sigma_{ij}(\mu)}{\partial a(\mu)} 
\underbrace{\frac{\partial a(\mu)}{\partial\log(\mu^2)}}_{\beta(\mu)a(\mu)} = \ord(a^5) \, .
\end{equation}
If there are more scale dependent quantities entering the cross section, such as running $\overline{\text{MS}}$-masses, their scale translations have to be included as well. The full
dependence on the scale $\mu$ will, by construction, be of the next order in $a$ once again.

%%%%%%%%%%%%%%%%%%%%%%%%%%%%%%%%%%%%%%%%%%%%%%%%%%
%%%%% Cross section and splitting kernels
%%%%%%%%%%%%%%%%%%%%%%%%%%%%%%%%%%%%%%%%%%%%%%%%%%
\section{Ingredients}
\label{sec:ingredients}
From the previous section, we conclude that we need the following ingredients to obtain all convolutions required for the \n3lo gluon fusion cross section:
\begin{itemize}
  \item The LO partonic cross section through $\ord(\epsilon^3)$.
  \item The NLO partonic cross section through $\ord(\epsilon^2)$.
  \item The NNLO partonic cross section through $\ord(\epsilon)$.
  \item The LO splitting kernels $P_{gg}^{(0)}$, $P_{gq}^{(0)}$, $P_{qg}^{(0)}$, $P_{qq}^{(0)}$.
  \item The NLO splitting kernels $P_{gg}^{(1)}$, $P_{gq}^{(1)}$, $P_{qg}^{(1)}$, $P_{qq}^{(1)}$, $P_{q\bar{q}}^{(1)}$, $P_{qQ}^{(1)}$ (where $q\neq Q \neq \bar{q}$).
  \item The NNLO splitting kernels $P_{gg}^{(2)}$ and $P_{gq}^{(2)}$ (owing to the fact that at LO, only the $gg$-channel is nonzero).
\end{itemize}
\subsection{Partonic cross section}
We work in the effective five-flavour theory with the top quark integrated out. This approximation has been shown to be very good (less than $5\%$) for light Higgs masses, as can be seen
by comparing the NLO results in effective and full six-flavour theory and by studying the importance of $1/m_t$ corrections of the effective NNLO cross section~\cite{Pak:2009dg,Harlander:2009mq}.
We expect this behaviour to persist at \n3lo.

The effective Lagrangian describing the interaction between gluons and the Higgs boson is given by
\begin{equation}
\mathcal{L}_{\text{eff}} = -\frac{1}{4v}C_1^{(B)}\,G^a_{\mu\nu} G^{a\mu\nu} H \, ,
\end{equation}
where $G^a_{\mu\nu}$ denotes the gluonic field-strength tensor. The Wilson coefficient $C_1$ which starts at $\ord(a)$ has been computed perturbatively to 
four-loop accuracy \cite{Chetyrkin:1997un,Schroder:2005hy} in the SM
as well as to three-loop accuracy for some BSM models~\cite{Anastasiou:2010bt,Furlan:2011uq,Pak:2010cu}. Through $\ord(a^4)$, the Wilson coefficient in the SM reads
\begin{align}
C_1 =& -\frac{a(\mu)}{3}\Bigg\{1 +a(\mu)\,\frac{11}{4} + a^2(\mu) \left[\frac{2777}{288}+\frac{19}{16}L_t+N_F\left(-\frac{67}{96}+\frac{1}{3}L_t\right)\right]\nn\\
  &+ a^3(\mu)\Bigg[\left(-\frac{6865}{31104}+\frac{77}{1728}L_t-\frac{1}{18}L_t^2\right)N_F^2 \nn\\
  & + \left(\frac{23}{32}L_t^2+\frac{55}{54}L_t+\frac{40291}{20736}-\frac{110779}{13824}\zeta_3\right)N_F \nn\\
  &  -\frac{2892659}{41472}+\frac{897943}{9216}\zeta_3+\frac{209}{64}L_t^2+\frac{1733}{288}L_t\Bigg] +\ord(a^4) \Bigg\} \, ,
\end{align}
with $L_t = \log(\mu^2/m_t^2)$. $N_F$ denotes the number of light flavours  set to 5.\\
The above expression denotes the renormalised Wilson coefficient, which is related to the bare one through the renormalisation constant,
\begin{equation}
C_1^{(B)} = Z_1(\mu)\,C_1(\mu) \, ,
\end{equation}
with
\begin{equation}
Z_1(\mu) = \frac{1}{1-\frac{\beta(\mu)}{\epsilon}} = 1-a\frac{\beta_0}{\epsilon}+a^2\left(\frac{\beta_0^2}{\epsilon^2}-\frac{\beta_1}{\epsilon}\right) 
+a^3\left( -\frac{\beta_0^3}{\epsilon^3} + 2\frac{\beta_0\beta_1}{\epsilon^2} - \frac{\beta_2}{\epsilon} \right),
\end{equation}
where we have suppressed the scale dependence of the strong coupling constant.

The partonic cross section for the production of a Higgs boson through gluon fusion can then be cast in the form%\footnote{Note that we have absorbed the factor of $1/z$, see eq.~\ref{sigma_hat_definition}, into the $\tilde{\sigma}^{(n,m)}_{ij}$, as well.}
\begin{equation}
\sigma_{ij} (z) = C_1^2\sigma_0\, \sum_{n,m=0}^\infty \tilde{\sigma}_{ij}^{(n,m)}(z)a^n \epsilon^m
\end{equation}
where we kept the squared Wilson coefficient factorised and pulled out all dimensionful prefactors, such that the $\eps^0$-piece of the leading order cross section becomes just
\begin{equation}
\tilde{\sigma}^{(0,0)}_{ij}(z) = \delta_{ig}\delta_{jg}\delta(1-z) \, .
\end{equation}
 All convolutions calculated in this work are done using the $\tilde{\sigma}^{(n,m)}_{ij}$ and from
here on, the term ``cross section'' will refer to these objects.

The sole dependence of  the LO cross section on $\epsilon$ is an overall factor of\\
$(1-\epsilon)^{-1} = \sum_{n=0}^\infty\epsilon^n$ from averaging over the $D$-dimensional polarisations of the
initial gluons. Thus, the LO partonic cross section through $\ord(\epsilon^3)$ is trivially found to be
\begin{equation}
\tilde{\sigma}^{(0,m)}_{ij}(z) = \delta_{ig}\delta_{jg}\delta(1-z) \, ,
\end{equation}
for all $m=0,\ldots,3$.

At NLO, the dependence on $\epsilon$ is still fairly simple. There are only two master integrals and they are easily computed to all orders in $\epsilon$.

The NNLO cross section through $\ord(\epsilon)$ necessitated the knowledge of the 29 master integrals to sufficiently high order in $\epsilon$. The double-virtual master integrals can be
found in work on the two-loop gluon form factor~\cite{Gonsalves:1983nq,Kramer:1986sr,Gehrmann:2005pd}. The real-virtual and double-real master integrals were
computed by two groups independently during the last year~\cite{Pak:2011hs,Anastasiou:2012kq}. The expression for the bare NNLO cross section in terms of master integrals was kindly provided to us by an author of~\cite{Anastasiou:2002yz}.

In general, the partonic cross sections consist of three types of terms, \emph{delta-, plus- and regular terms}.
\begin{equation}
\tilde{\sigma}_{ij}^{(n,m)}(x)  =  a_{ij}^{(n,m)}\,\delta(1-x) + \sum_k b_{ij}^{(n,m),k}\, \DD_k(1-x) + c_{ij}^{(n,m)}(x) \, ,
\label{eq:deltaplusreg}
\end{equation}
where the plus-distribution $\DD_k(1-x) = \left[ \frac{\log^k(1-x)}{1-x} \right]_+$ is defined via its action on a test function $f(x)$ with a finite value at $x=1$,
\begin{equation}
\int_0^1\mathrm{d}x\,\DD_k(1-x)\,f(x) = \int_0^1\frac{\log^k(1-x)}{1-x}\left(f(x)-f(1)\right) \, .
\end{equation}

The full expressions for the partonic cross sections through NNLO can be found in the ancillary files accompanying this arXiv publication. They agree with the ones given in~\cite{Hoschele:2012xc} after
compensating for the factor of $1/z$ that was not included in that publication.
\subsection{Splitting kernels}
\label{sec:AP}
The splitting kernel
\begin{equation}
P_{ij}(x) = \sum_{n=0}^\infty a^{n+1}\,P_{ij}^{(n)}(x)
\end{equation}
describes the probability of a parton $j$ emitting a collinear parton $i$ carrying a fraction $x$ of the momentum of the initial parton. The splitting kernels are known up to three loops ($P^{(2)}_{ij}$) and may all be found in~\cite{Vogt:2004mw,Moch:2004pa}.

Note some different conventions that we use, though. Since we chose to expand all our results in $a = \frac{\alpha_s}{\pi}$ as opposed to $\frac{\alpha_s}{4\pi}$ as in~\cite{Vogt:2004mw,Moch:2004pa},
our kernels $P^{(n)}_{ij}$ differ from the ones in the reference by a factor of $\left(\frac{1}{4}\right)^{n+1}$.

Also, since by $P^{(n)}_{qg}$ we mean the emission of a single quark of a given flavour, we differ from the expression in~\cite{Vogt:2004mw,Moch:2004pa} which parametrises the emission of \emph{any}
quark, by a factor of $\frac{1}{2N_F}$.

Furthermore, there is also a conventional difference to the splitting kernels used in~\cite{Hoschele:2012xc}. The authors of that publication use the quark-quark splitting kernel as
defined in eq. (2.4) of~\cite{Vogt:2004mw}. This kernel, which we shall denote by $\tilde{P}_{qq}$ is used in the DGLAP evolution of pdfs.
% but does not suffice 
To compute all contributions to the \n3lo gluon fusion cross section, we have to distinguish different initial-state channels such as $q\bar{q}$
(quark-antiquark), $qq$ (identical quarks) and $qQ$ (quarks of different flavour)
which are convoluted with different combinations of pdfs. Thus, for channel-by-channel collinear factorisation, we require the three distinct, flavour-dependent quark-quark kernels
\begin{equation}
P_{qq} \, , \quad P_{q\bar{q}} \, , \quad P_{qQ} \, ,
\end{equation}
which describe the emission of an identical quark, the emission of an antiquark of the same flavour, and the emission of a quark or antiquark of a different flavour, respectively.
The latter two kernels vanish at the one-loop order, $P_{q\bar{q}}^{(0)} = 0 = P_{qQ}^{(0)}$ but are nonzero for higher orders.
In the notation of~\cite{Moch:2004pa}, this corresponds to the kernels $P_{q_iq_j}$ and $P_{q_i\bar{q}_j}$. The relation between $\tilde{P}_{qq}$ and our kernels is given by
\begin{equation}
\label{eq:steinhausermap}
\tilde{P}_{qq} = P_{qq} + P_{q\bar{q}} +2(N_F-1)\,P_{qQ} \, .
\end{equation}
We are not aware of results in~\cite{Hoschele:2012xc} that involve the flavour-dependent quark-quark kernels.

We close this section by giving the expressions for the four LO splitting kernels. For the lengthy higher-order kernels, we again refer to the machine-readable files accompanying this publication. The two NNLO kernels were taken from~\cite{Vogtwebsite} in \verb+Form+ format and then translated to {\sc Maple} input. Their regular parts were tested against the \verb+Fortran+ routine in~\cite{Vogtwebsite}, and their $\delta(1-z)$ and $\DD_n(1-z)$ parts were checked against~\cite{Vogt:2004mw}.

\begin{eqnarray}
P^{(0)}_{gg}(x) &=& \beta_0\,\delta(1-x) + 3\left(\DD_0(1-x)+x(1-x)-2+\frac{1}{x}\right) \, . \\
P^{(0)}_{gq}(x) &=&  \frac{2}{3}\frac{1+(1-x)^2}{x}\, . \\
P^{(0)}_{qg}(x) &=&  \frac{1}{4}\left(x^2+(1-x)^2\right)\, . \\
P^{(0)}_{qq}(x) &=&  \delta(1-x)+\frac{2}{3}\left(2\DD_0(1-x)-1-x\right) \, .
\end{eqnarray}
%
%%%%%%%%%%%%%%%%%%%%%%%%%%%%%%%%%%%%%%%%%%%%%%%%%
%%%%% Convolutions
%%%%%%%%%%%%%%%%%%%%%%%%%%%%%%%%%%%%%%%%%%%%%%%%%%
\section{Computation of the convolutions}
\label{sec:conv}
In this section we will describe the method we used to compute the convolutions of splitting kernels and partonic cross sections that are needed to cancel collinear divergences at \n3lo.
Let us remark that our method is different from the technique used in \cite{Hoschele:2012xc}, where the convolutions were calculated in Mellin space (where convolutions turn into ordinary multiplications)
and the problem was essentially the calculation of the inverse Mellin-transform.

In the following, we restrict ourselves to a single convolution. Since the convolution product is associative, any multiple convolution appearing in the \n3lo cross section
can be obtained by repeating the steps using the result of the first convolution and the next convolutant. As already mentioned, both the splitting kernels  and the partonic cross sections
consist of three types of terms, \emph{delta-, plus- and regular terms}.
\begin{eqnarray}
\label{eq:sigmaexp}
\tilde{\sigma}_{ij}^{(n,m)}(x) & = & a_{ij}^{(n,m)}\,\delta(1-x) + \sum_k b_{ij}^{(n,m),k}\, \DD_k(1-x) + c_{ij}^{(n,m)}(x) \, ,\\
P_{ij}^{(n)}(x) & = & A_{ij}^{(n)}\,\delta(1-x) + \sum_k B_{ij}^{(n),k}\, \DD_k(1-x) + C_{ij}^{(n)}(x) \, ,
\end{eqnarray}
where the regular pieces $c_{ij}^{(n,m)}(x)$ and $C_{ij}^{(n)}(x)$ consist of harmonic polylogarithms (HPLs)
 times polynomials in $x$ or factors of $1/x$ and $1/(1-x)$. Harmonic polylogarithms are a generalisation of the ordinary logarithm and the classical polylogarithms,
\begin{equation}
  \log(x) = \int_1^z\frac{\dd t}{t} \, \quad \text{and} \quad \mathrm{Li}_n(z) = \int_0^z\frac{\dd t}{t}\,\mathrm{Li}_{n-1}(t) \, ,
\end{equation}
where $\mathrm{Li}_1(z) = -\log(1-z)$. HPLs can be defined recursively via the integral
\begin{equation}
  \mathrm{H}(a_1,a_2,\ldots,a_n;z) = \int_0^z\dd t f_{a_1}(t)\,\mathrm{H}(a_2,\ldots,a_n;t) \, , \quad a_i\in\{-1,0,1\} \, ,
\end{equation}
with
\begin{equation}
  f_{-1}(x) = \frac{1}{1+x} \, , \quad f_0(x) = \frac{1}{x} \, , \quad f_1(x) = \frac{1}{1-x} \, ,
\end{equation}
and in the special case where all $a_i=0$, the HPL is defined as
\begin{equation}
  \mathrm{H}(\vec{0}_n;z) = \frac{1}{n!}\log^n(z) \, .
\end{equation}
For more comprehensive information about harmonic polylogarithms, we refer to~\cite{Remiddi:1999ew,Gehrmann:2001pz,Vollinga:2004sn,Maitre:2005uu,Maitre:2007kp,Duhr:2011zq,Buehler:2011ev}.

Any convolution involving a delta function trivially returns the other convolutant (whether it be another delta function, a plus distribution or a regular function),
\begin{equation}
\big(\delta\otimes f\big)(z) = \int_0^1\mathrm{d}x\mathrm{d}y\, \delta(1-x)f(y)\delta(xy-z) = \int_0^1\mathrm{d}y\, f(y)\delta(y-z) = f(z) \, .
\end{equation}
Convolutions involving two plus-distributions are more involved, yet no integral actually has to be solved. We comment on their calculation and list results for the
required plus-plus convolutions in appendix \ref{app:plusplusconv}.

For the remaining two types of convolutions, we end up with an actual integral that we need to compute,
\begin{eqnarray}
\big(\DD_n\otimes f\big) (z) & = & \int_0^1\mathrm{d}x\mathrm{d}y\DD_n(1-x)f(y)\delta(xy-z) = \int_z^1\mathrm{d}x\DD_n(1-x)\frac{f\left(\frac{z}{x}\right)}{x} \nonumber \\
 &=& \frac{\log^{n+1}(1-z)}{n+1}f(z) + \int_z^1\mathrm{d}x\frac{\log^n(1-x)}{1-x}\left(\frac{f\left(\frac{z}{x}\right)}{x} - f(z)\right)\, ,\\
\big(f\otimes g\big) (z) & = & \int_0^1\mathrm{d}x\mathrm{d}y f(x)g(y)\delta(xy-z) = \int_z^1\mathrm{d}x f(x)g\left(\frac{z}{x}\right)\frac{1}{x} \, .
\end{eqnarray}
Note the boundary term we pick up when evaluating a plus-distribution in an interval which is different from $(0,1)$.

Even though we know, from the denominator-structures appearing in the integrals, that our final result can only consist of HPLs
(times polynomials in $z$ and factors of $\frac{1}{z}$ or $\frac{1}{1-z}$), we have to step into 
the realm of multiple polylogarithms (MPLs) in intermediate steps of the calculation. MPLs are defined analogously to HPLs, but allow for any complex number in the index vector
instead of only $\{-1,0,1\}$ in the HPL case. Recursively,
\begin{equation}
\label{eq:MPLintegraldef}
G(x_1,x_2,\ldots,x_n;z) = \int^z_0\mathrm{d}t \frac{G(x_2,\ldots,x_n;t)}{t-x_1} \, , \,  \{x_i\}\in\mathbb{C} \quad \text{and} \quad G(z) = 1 \, .
\end{equation}
Specifically, a MPL may be a function of multiple variables that appear anywhere in the index vector $(x_1,\ldots,x_n)$. The relation to HPLs reads
\begin{equation}
\mathrm{H}(a_1,\ldots,a_n;x) = (-1)^k G(a_1,\ldots,a_n;x) \, , \quad \{a_i\}\in\{-1,0,1\} \, ,
\end{equation}
where $k$ is the number of $+1$ indices in $(a_1,\ldots,a_n)$. This sign difference is due to the fact that HPLs historically use $\frac{1}{1-t}$ as the weight function when adding a $+1$
to the index vector.

For more detailed information on multiple polylogarithms, see references~\cite{Goncharov:1998,Goncharov:2001,Duhr:2012fh} and references therein. Note that
the order of the MPLs indices is often reversed. We follow the convention of~\cite{Duhr:2012fh}.\\
The subsequent steps to solve the integrals are as follows:
\begin{enumerate}
\item We first remap the integral by $x\mapsto 1-x$, such that the integration region becomes $(0,1-z)$.
\item HPLs with argument $1-x$ and $\frac{z}{1-x}$ have to be written as a combination of MPLs with the integration variable $x$ as their argument, or no $x$-dependence at all.
For example,
\begin{eqnarray}
\mathrm{H}\left(1;\frac{z}{1-x}\right) & = & -\log\left( 1-\frac{z}{1-x}\right) = -\log\left(\frac{1-x-z}{1-x}\right) \nonumber \\
 & = & \log(1-x) - \log\left((1-z)\left(1-\frac{x}{1-z}\right)\right) \nonumber \\
 & = & G(1;x) - G(1;z)-G(1-z;x) \, ,
\end{eqnarray}
where we've used that $G(a;b) = \log\left(1-\frac{b}{a}\right)$ for $a\neq0$. For MPLs of higher weights, one can find these translations by using the recursive definition of MPLs and changing
variables in the integration. This becomes very tedious, though, so it proved to be more practical to use the symbol formalism developed in recent 
years~\cite{Goncharov-simple-Grassmannian,Goncharov:2010jf,Duhr:2011zq,Duhr:2012fh} and to follow the method presented in appendix D of~\cite{Anastasiou:2013srw}.
For technical details, we refer the reader to said appendix.
\item We are left with integrals of a single MPL with argument $x$ times factors of $x^k$ ($k\geq -1$) or $\frac{1}{1-x}$, which can all immediately be solved via
  the MPLs recursive definition and integration by parts.
\item At this stage, all integrations have been performed. The result still contains MPLs where the variable $z$ appears multiple times in the argument vector. Using
  the techniques from appendix D of~\cite{Anastasiou:2013srw} again, we can rewrite all the expressions in HPLs.

\item The final numerical check on the result consists of the comparison of the original integral using Mathematica's numerical integration (using the package \verb+HPL+~\cite{Maitre:2005uu,Maitre:2007kp}
to evaluate the HPLs numerically) and our final expression, using \verb+Ginsh+, the interactive frontend of the computer algebra system GiNaC~\cite{Vollinga:2004sn}, for a random value of $z$.
\end{enumerate}
The full set of convolutions can be found in machine-readable form (both  {\sc Maple} and {\sc Mathematica}) in the ancillary files accompanying this arXiv publication.
They were all compared analytically in {\sc Mathematica} to the expressions given in~\cite{kitwebsite}, and complete agreement was found for all convolutions. For convolutions
involving the two-loop quark-quark splitting kernels $P^{(1)}_{qq}$, $P^{(1)}_{q\bar{q}}$ and $P^{(1)}_{qQ}$, the results had to be combined according to eq.~(\ref{eq:steinhausermap}) to
find equality.
\subsection{Soft expansion of the convolutions}
\label{sec:softlimits}
While the full \n3lo corrections to the gluon fusion cross section may still be out of reach for the time being, a description in the soft limit could be feasible already in the close future.
Note that this was the sucession at NNLO, as well, where the expansion of the cross section up to $\ord((1-z)^{16})$~\cite{Harlander:2002wh,Catani:2001ic} was published before the full 
computation~\cite{Anastasiou:2002yz,Ravindran:2003um}.
The numerical agreement between the two computations proved to be excellent, so, anticipating the same behaviour at \n3lo, the soft expansion of the \n3lo corrections
would be a very important result to obtain. The first pure \n3lo piece of the third order soft expansion, the soft triple-real emission contribution, has recently been published~\cite{Anastasiou:2013srw}.

In the limit $z\rightarrow 1$, the partonic cross section (and all convolutions contributing to it) can be cast in the following form (suppressing partonic indices)
\begin{equation}
\tilde{\sigma}^{(n,m)}(z) = a^{(n,m)}\delta(1-z) + \sum_{k=0}^{2n-1}b^{(n,m),k} \DD_k(1-z) + \sum_{k=0}^{2n-1}\sum_{l=0}^\infty c_{kl}^{(n,m)} \, \log(1-z)^k\,(1-z)^l \, .
\end{equation}
We thus need to expand the regular part as a polynomial in $(1-z)$, times $\log(1-z)$ terms. We proceed as follows:
\begin{enumerate}
\item We define $z'\equiv 1-z$. Thus, our expressions now consist of HPLs with argument $1-z'$ times powers of $z'$. The desired limit is $z'\rightarrow 0$.
\item We want to rewrite the HPLs with argument $1-z'$ as MPLs with argument $z'$, which results in changing the array of indices from $\{-1,0,1\}$ to $\{0,1,2\}$, as can be easily seen
by taking the integral definition eq.~\eqref{eq:MPLintegraldef} for $x_1\in\{-1,0,1\}$ and changing variables $t\mapsto 1-t$. The rewriting is achieved once again with
the techniques from appendix D of~\cite{Anastasiou:2013srw}.
\item The expansion of any MPL in its argument is straightforward, since there is a connection between MPLs and multiple nested sums~\cite{Goncharov:1998},
\begin{equation}
\label{eq:nestedsums}
\mathrm{Li}_{m_1,\ldots,m_k}(x_1,\ldots,x_k) = \sum_{n_k=1}^\infty \frac{x_k^{n_k}}{n_k^{m_k}}\sum_{n_{k-1}=1}^{n_k-1}\cdots\sum_{n_{1}=1}^{n_2-1} \frac{x_1^{n_1}}{n_1^{m_1}} \, 
\end{equation}
where the translation from MPLs to nested sums is given by
\begin{align}
%\mathrm{Li}_{m_1,\ldots,m_k}(x_1,\ldots,x_k) =(-1)^kG\left( \underbrace{0,\ldots,0}_{m_k-1},\frac{1}{x_k},\ldots,\underbrace{0,\ldots,0}_{m_1-1},\frac{1}{x_1\cdots x_k};1 \right) \, .
&G\Big( \underbrace{0,\ldots,0}_{m_k-1},1,\underbrace{0,\ldots,0}_{m_{k-1}-1},a_{k-1},\ldots,\underbrace{0,\ldots,0}_{m_1-1},a_1;x_k \Big) \nn\\
&=(-1)^{k}\,\mathrm{Li}_{m_1,\ldots,m_k}\left(\frac{a_2}{a_1},\frac{a_3}{a_2},\ldots,\frac{a_{k-1}}{a_{k-2}},\frac{1}{a_{k-1}},x_k\right)
\end{align}
The specific form of the MPL on the left-hand side of the equation above can be obtained via the scaling property, $G(x_1,\ldots,x_n;z) = G(\lambda x_1,\ldots,\lambda x_n,\lambda z)$, where 
$\lambda\neq 0\neq x_n$. MPLs with a rightmost index of $0$ must be rewritten using the shuffle product, e.g.
\begin{equation}
G(a,0;x) = G(0;x)G(a;x) - G(0,a;x) = \log(x) G(a;x) - G(0,a;x) \, ,
\end{equation}
until all rightmost zeroes have been turned into explicit logarithms. The remaining MPLs can then be safely translated to nested sums.

The crucial point is that the variable $x_k$ only appears in the outermost sum in eq.~\eqref{eq:nestedsums}, while the inner nested sums only depend on the $x_{i<k}$,
which in our case are the indices $a_i\in\{0,1,2\}$. We thus easily obtain the desired expansion when we just truncate the sum over $n_k$ in eq.~\eqref{eq:nestedsums} 
at the highest power of $z'$ we are interested in.
\item The validity of the soft expansions of the convolutions was checked numerically for some small values of $z'$. All soft expansions of the convolutions
up to $\ord\left((1-z)^{12}\right)$ can be found in the ancillary files accompanying this arXiv publication.
\end{enumerate}
%
%%%%%%%%%%%%%%%%%%%%%%%%%%%%%%%%%%%%%%%%%%%%%%%%%
%%%%% Results
%%%%%%%%%%%%%%%%%%%%%%%%%%%%%%%%%%%%%%%%%%%%%%%%%%
\section{Numerical results for the gluon fusion scale variation at \n3lo}
\label{sec:results}
The total cross section for Higgs production through gluon fusion at N$^3$LO depends on the factorisation and renormalisation scales explicitly, through logarithmic terms that have been derived in this work, and implicitly through the $\mu_r$ dependence of $\alpha_s$ and the $\mu_f$ dependence of the parton distribution functions. In principle N$^3$LO parton distribution functions should be used, but in practice, not only  are they not available (nor will they be in the near future), but also their deviation with respect to the NNLO pdfs available is expected to be very small. On the other hand, the full, implicit, $\mu_r$ dependence through $\alpha_s$, can only be estimated once the N$^3$LO matrix elements are known, and in particular the coefficients\footnote{Note that we define the coefficients $b_{ij}$ of the plus terms to be independent of $z$. This implies that the regular terms $c_{ij}^{(n,m)}(z)$ contain terms with the logarithms $\log(z)$ and $\log(1-z)$.} $a_{ij}^{(3,0)}$, $b_{ij}^{(3,0),k}$ and $c_{ij}^{(3,0)}(z)$ in eq.~\eqref{eq:deltaplusreg}.  Of these contributions only the $b_{ij}^{(3,0),k}$ are known, from mass factorisation constraints~\cite{Moch:2005ky}. The question, then, arising is whether we can anticipate the scale uncertainty at N$^3$LO with the information currently available. 

To this end we parametrize the unknown delta and regular coefficients by a scaling factor $K$ times the corresponding NNLO coefficients:
\begin{equation}
a_{ij}^{(3,0)} = K\,a_{ij}^{(2,0)} \, , \qquad  f_i \otimes f_j \otimes c_{ij}^{(3,0)}(z) = K\left(f_i\otimes f_j \otimes c_{ij}^{(2,0)}(z)\right) \, ,
\end{equation}
There is no a priori reason why the scaling factor for the delta and the regular terms should be the same. However, it turns out that the numerical impact of the delta coefficient $a_{ij}^{(3,0)}$ is negligible (for scaling coefficients that do not break by orders of magnitude the pattern observed from lower orders), in contrast with the coefficient of the regular part, so we adopt here a common scaling factor to keep the parametrisation simple. For the same reason we use the same $K$ scaling coefficient for all initial state channels.   
\begin{center}
\FIGURE{
\includegraphics[width=1.0\textwidth]{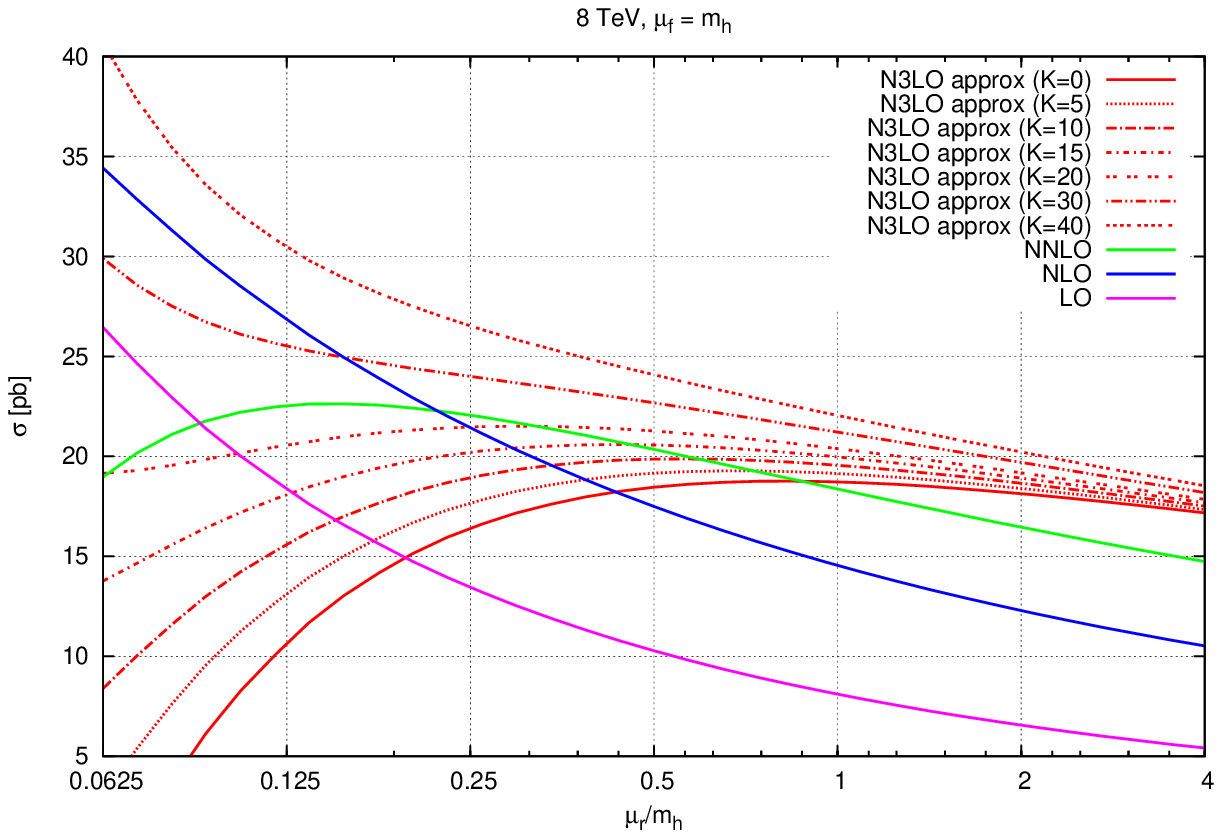}
\caption{Scale variation of the different orders of the gluon fusion cross section at 8 TeV. $\mu_f$ is fixed to $m_h$ and only $\mu_r$ is varied. The scaling coefficient $K$ is varied from 0 to 40 to estimate the impact of the unknown \n3lo contributions.}
\label{fig:muronly8tev}
}
\end{center}

A loose argument about the size of $K$ can be derived if one assumes a good perturbative behaviour at $\mu_r=\mu_f=m_H$ where all other terms of order $a^5$ vanish. Since $a(m_H)\sim 1/30$ one expects $K$ not to be much larger than $30$. For comparison, the corresponding rescaling factors between NNLO and NLO are 
\begin{equation}
\frac{f_g\otimes f_g \otimes c_{gg}^{(2,0)}}{f_g \otimes f_g \otimes c_{gg}^{(1,0)}} \sim 30 \, , \quad \frac{a_{gg}^{(2,0)}}{a_{gg}^{(1,0)}} \sim 1.5 \, ,
\end{equation}
for $m_h = 125$ GeV and $\mu_f =\mu_r= m_h$.
\begin{center}
\FIGURE{
\includegraphics[width=1.0\textwidth]{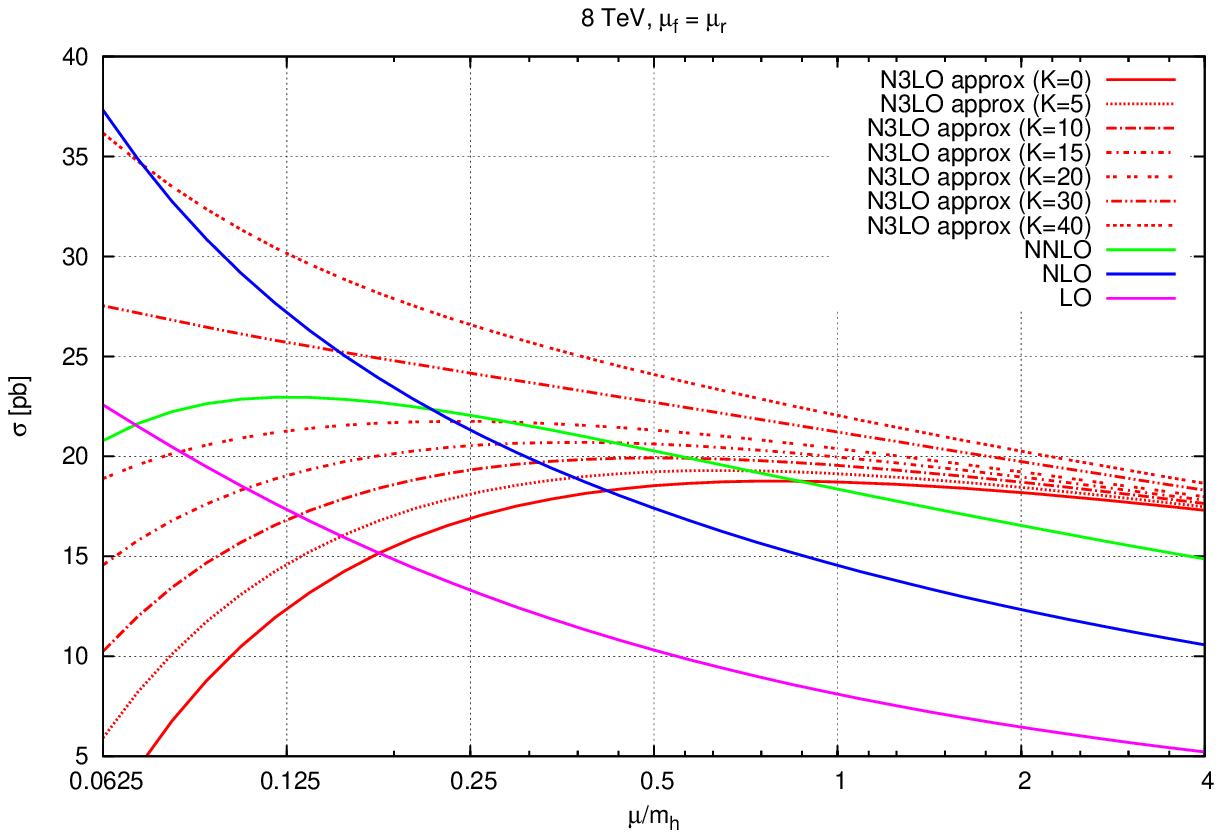}
\caption{Scale variation of the different orders of the gluon fusion cross section at 8 TeV. $\mu_f$ and $\mu_r$ are varied simultaneously. The scaling coefficient $K$ is varied from 0 to 40 to estimate the impact of the unknown \n3lo contributions.}
\label{fig:mufeqmur8tev}
}
\end{center}
 In what follows we study the inclusive cross section as a function of the scales, in the HQET approach, rescaled with the exact leading order cross section. We use the framework of the {\tt iHixs}
program~\cite{Anastasiou:2011pi} a \verb+Fortran+ code which contains the complete NNLO cross section for gluon fusion in HQET. The coupling $\alpha_s$ was run to four-loop order according to eq.~\eqref{eq:rge}, while for the parton distributions, the MSTW08 NNLO set was used. Furthermore, to cross-check our results, a second implementation was programmed in \verb1C++1, where the convolutions of splitting kernels and partonic cross sections were performed numerically. For both codes, the numerical evaluation of HPLs was performed using the library {\sc Chaplin}~\cite{Buehler:2011ev}. The two implementations agreed for all parameter configurations that were tested.

\begin{figure}
  \centering
  \begin{subfigure}[b]{0.6\textwidth}
 %   \centering
    \hspace*{-0.15\textwidth}
    \includegraphics[width=\textwidth]{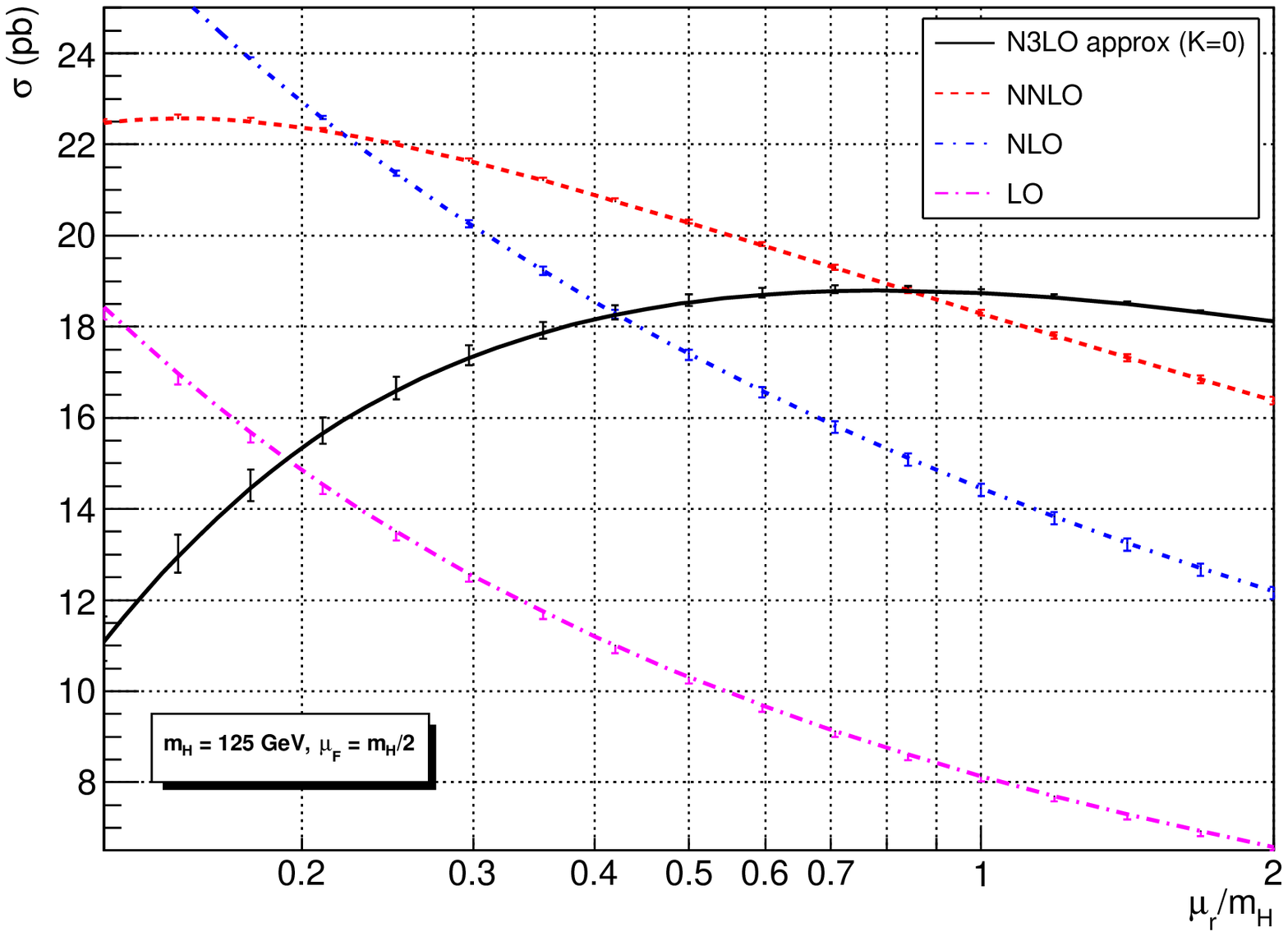}
    \caption{}
  \end{subfigure}%
  ~
  \begin{subfigure}[b]{0.6\textwidth}
    %\centering
    \hspace*{-0.27\textwidth}
    \includegraphics[width=\textwidth]{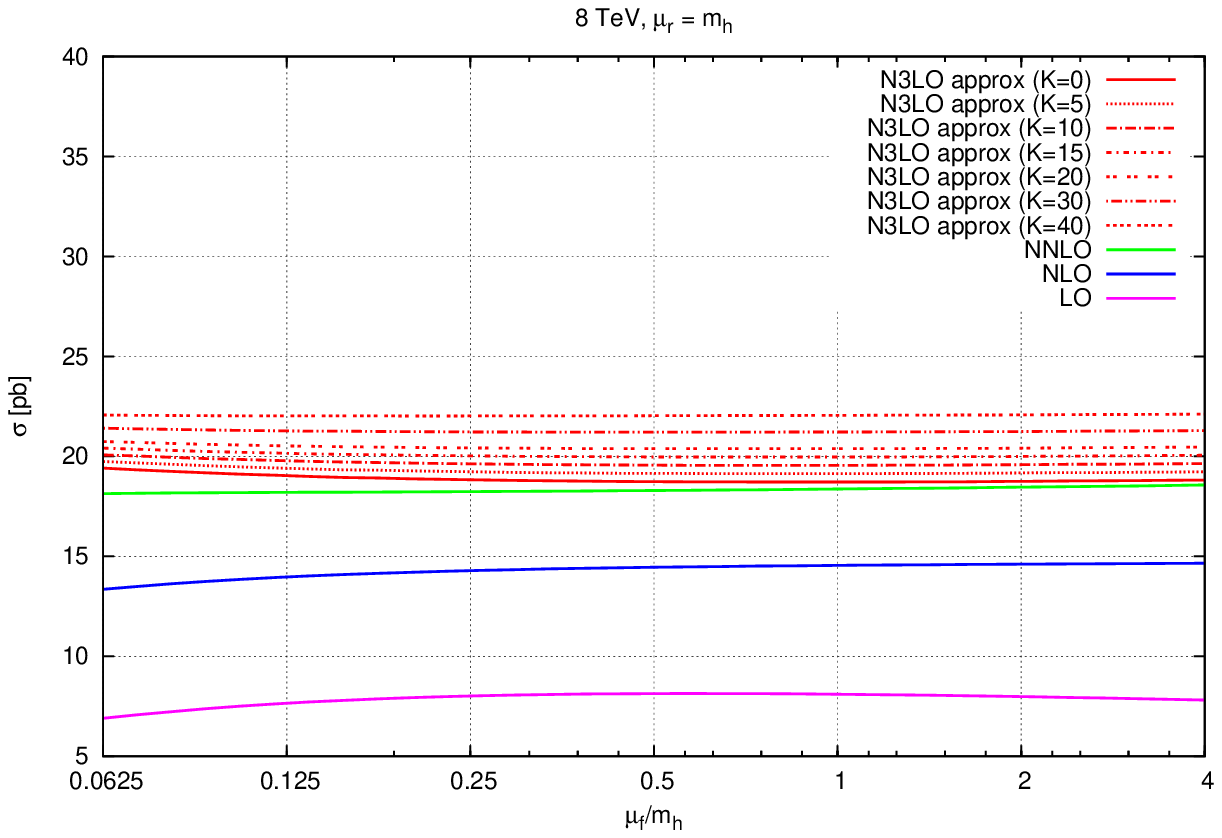}
    \caption{}
  \end{subfigure}
   \caption{Scale variation of the different orders of the gluon fusion cross section at 8 TeV. In (a) $\mu_r$ is varied along the $x$-axis, while the bars represent variation of $\mu_f$ around the central value $m_h/2$. In (b) $\mu_r$ is fixed to $m_h$ and only $\mu_f$ is varied. The scaling coefficient $K$ is varied from 0 to 40.}
  \label{fig:mufonly8tev}
\end{figure}

In figure~\ref{fig:muronly8tev}, the different orders of the hadronic gluon fusion cross section for the 8 TeV LHC and a Higgs mass of 125 GeV, along with several N$^3$LO approximants for various numerical values of $K$ are plotted as a function of the renormalisation
scale $\mu_r$, while the factorisation scale is fixed to $\mu_f=m_h$. Note that the convolutions of splitting kernels and partonic cross sections do not enter in this plot, since they are
proportional to $\log (\mu_f^2/m_h^2)$. The $\mu_r$ scale variation for LHC with 14 TeV centre-of-mass energy is shown in fig.~\ref{fig:muronly14tev}. 
The $\mu_f$ scale dependence, shown in figure~\ref{fig:mufonly8tev} for 8 TeV centre-of-mass energy, is, as expected, extremely mild, in accordance with what is observed at NNLO.
\begin{center}
\FIGURE{
\includegraphics[width=1.0\textwidth]{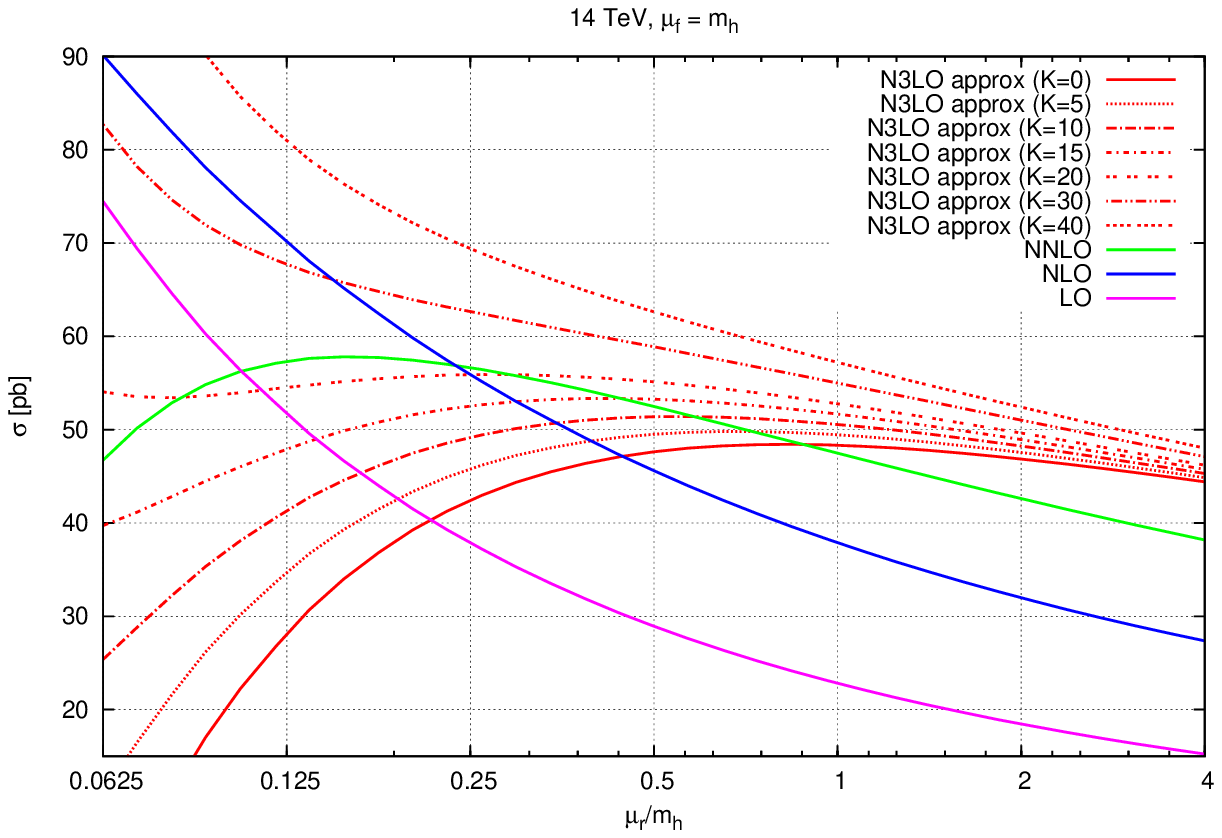}
\caption{Scale variation of the different orders of the gluon fusion cross section at 14 TeV. $\mu_f$ is fixed to $m_h$ and only $\mu_r$ is varied. The scaling coefficient $K$ is varied from 0 to 40.}
\label{fig:muronly14tev}
}
\end{center}
\begin{center}
\FIGURE{
\includegraphics[width=1.0\textwidth]{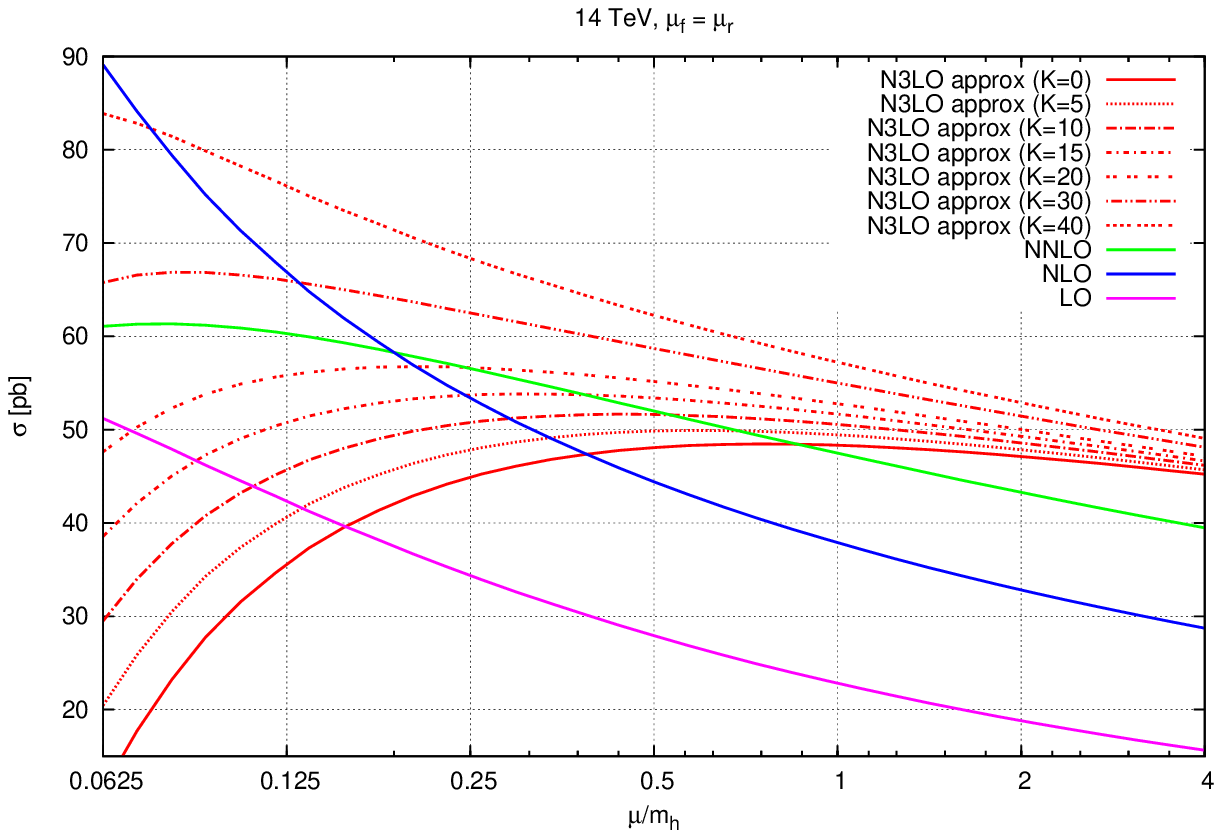}
\caption{Scale variation of the different orders of the gluon fusion cross section at 14 TeV. $\mu_f$ and $\mu_r$ are varied simultaneously. The scaling coefficient $K$ is varied from 0 to 40.}
\label{fig:mufeqmur14tev}
}
\end{center}

Figures~\ref{fig:mufeqmur8tev} and \ref{fig:mufeqmur14tev} display the overall scale dependence, with both scales set to be equal and varied simultaneously.
We note that the curves for the approximate \n3lo cross section with various $K$s spread widely in the low  scale region, i.e. for $\mu<30$ GeV. This is not unreasonable, though, as
in this regime, the unknown \n3lo contributions that are neglected  become much more important due to the running of $\alpha_s$. Indeed, at the lowest renormalisation scale considered,
$\mu=m_h/16\approx 7$ GeV, the coupling becomes as big as $\alpha_s \approx 0.2$, i.e. we are barely in the perturbative regime. The term
\begin{equation}
\tilde{\sigma}^{(3)}_{gg}(\mu) \quad\ni\quad 4\beta_0\,\log\left( \frac{\mu^2}{m_h^2} \right) \tilde{\sigma}^{(2)}_{gg}(m_h) 
\end{equation}
which is supposed to cancel the implicit logarithms in the running of $\alpha_s$, and which becomes large and negative, thus pulls the curve down for small scales, and is canceled by the currently unknown contributions whose magnitude is small at $\mu_r=m_h$ but is greatly enhanced due to $\alpha_s$ at small $\mu_r$.
It can hardly be overemphasised that the above prescription does not represent a proper calculation of the \n3lo matrix elements, but just a way of parametrising
their unknown numerical importance. Once the height of the \n3lo curve at $(\mu_r,\mu_f) = (1,1)$ is set, the shape of the full curve only depends
on lower order cross sections (which we know exactly), the running of $\alpha_s$ and the parton distribution functions, respectively.

As mentioned above, the unknown, numerically important coefficient functions $c_{gg}^{(3,0)}(z)$ contain logarithmic contributions that are singular at threshold, $\log(1-z)$, contributions that are regular and contributions that are singular at the opposite, high energy limit $\log(z)$. The leading and several, but not all,  subleading threshold contributions are associated with multiple soft emissions and can be recovered by resummation techniques. The authors of~\cite{Moch:2005ky} have used the expressions for $b_{ij}^{(3,0),k}$ that they have derived to perform a soft approximation in Mellin space, resulting in a N$^3$LO approximant with a scale uncertainty of $\approx 4\%$. Recently, the authors of~\cite{Ball:2013bra} estimate  $c_{gg}^{(3,0)}(z)$ by interpolating in Mellin space, between the soft approximation (that captures threshold logarithms) and the BFKL limit (that captures high energy $\log(z)$ terms). This approach matches the NNLO cross section neatly, and  results in an approximant for the \n3lo with a scale uncertainty of $7\%$ if the scale is varied in the interval $[m_h/2,2m_h]$ (or smaller, if the interval chosen is $[m_h/4,m_h]$).   

Indeed, by comparing our results for the $\mu_r$-dependence of the \n3lo cross section for the dominant gluon gluon initial state, with the numbers obtained via the recently released
numerical program \verb+gghiggs+~\cite{Ball:2013bra}, we find agreement between the two curves
when setting $K$ to 25, as is displayed in figure~\ref{fig:bonvini}.

\begin{center}
\FIGURE{
\includegraphics[width=1.0\textwidth]{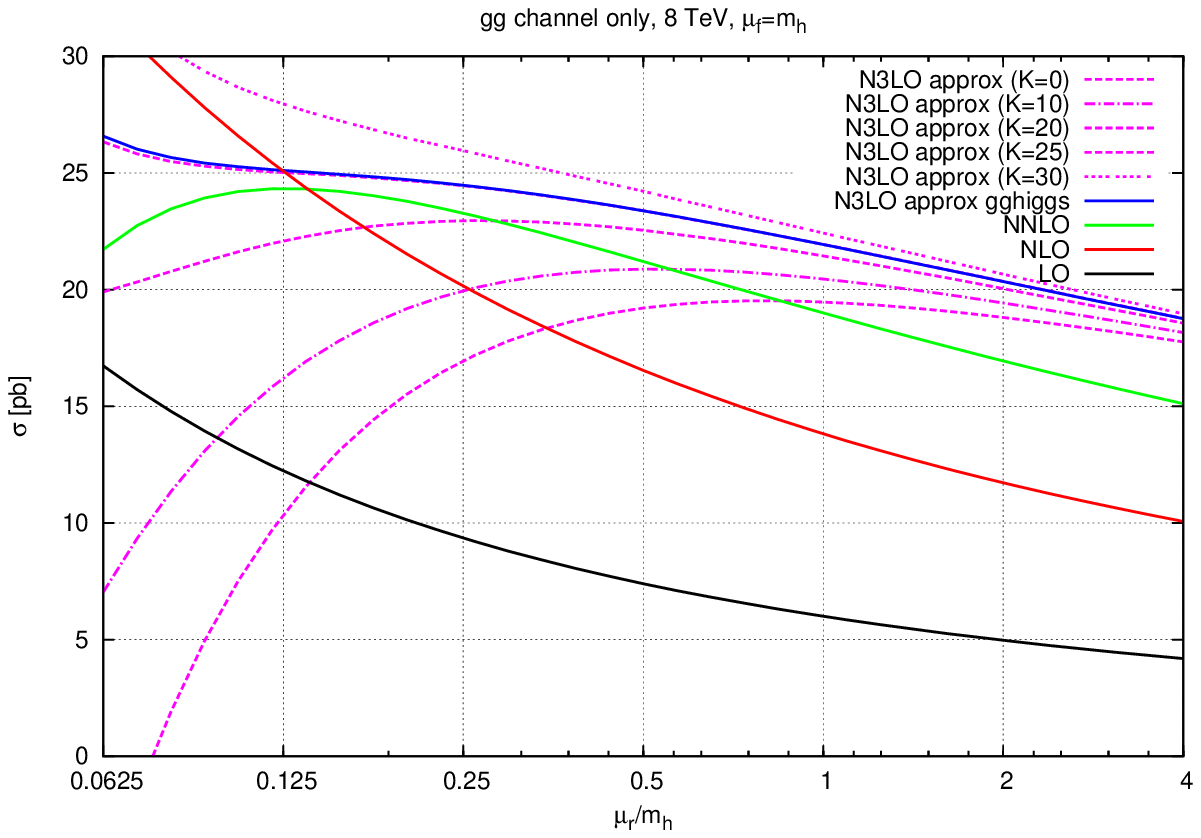}
\caption{Scale variation of the different orders of the gluon fusion cross section at 8 TeV. $\mu_f$ is fixed to $m_h$ and only $\mu_r$ is varied. $K$ is varied from 0 to 30. Only the $gg$ channel is plotted, and compared to the results obtained with~\cite{Ball:2013bra}.}
\label{fig:bonvini}
}
\end{center}

While it is plausible that the leading logarithmic contributions, being threshold enhanced, capture the bulk of the cross section, it is unclear whether the unknown subleading contributions, as well as the non-logarithmic terms, are really negligible. Their importance certainly rises for the LHC at 14 TeV, as the luminosity function suppresses the region away from threshold less, resulting in more phase space for real radiation.  
One might, therefore, want to be conservative about their magnitude, and hence on the size of the scale uncertainty to be anticipated before the full \n3lo result is available. Table~\ref{table:scaleuncertainty} shows the estimates for various values of the rescaling factor $K$, covering the range from relatively mild to extremely strong \n3lo corrections, resulting in scale uncertainties varying from $2\%$ to as large as $8\%$ or more. The scale uncertainties cited here are evaluated by varying the scales in the interval $[m_h /4, m_h]$. 

\begin{table}
\centering
\input{data/table_mstw}
\caption{Cross sections and scale uncertainties for the 8 TeV LHC. The central scale choice is $\mu_r = \mu_f = m_h/2$, and uncertainties are found by varying the two scales simultaneously by a factor of two.}
\label{table:scaleuncertainty}
\end{table}

\begin{figure}
  \centering
  \begin{subfigure}[b]{0.30\textwidth}
    \centering
    \includegraphics[width=\textwidth]{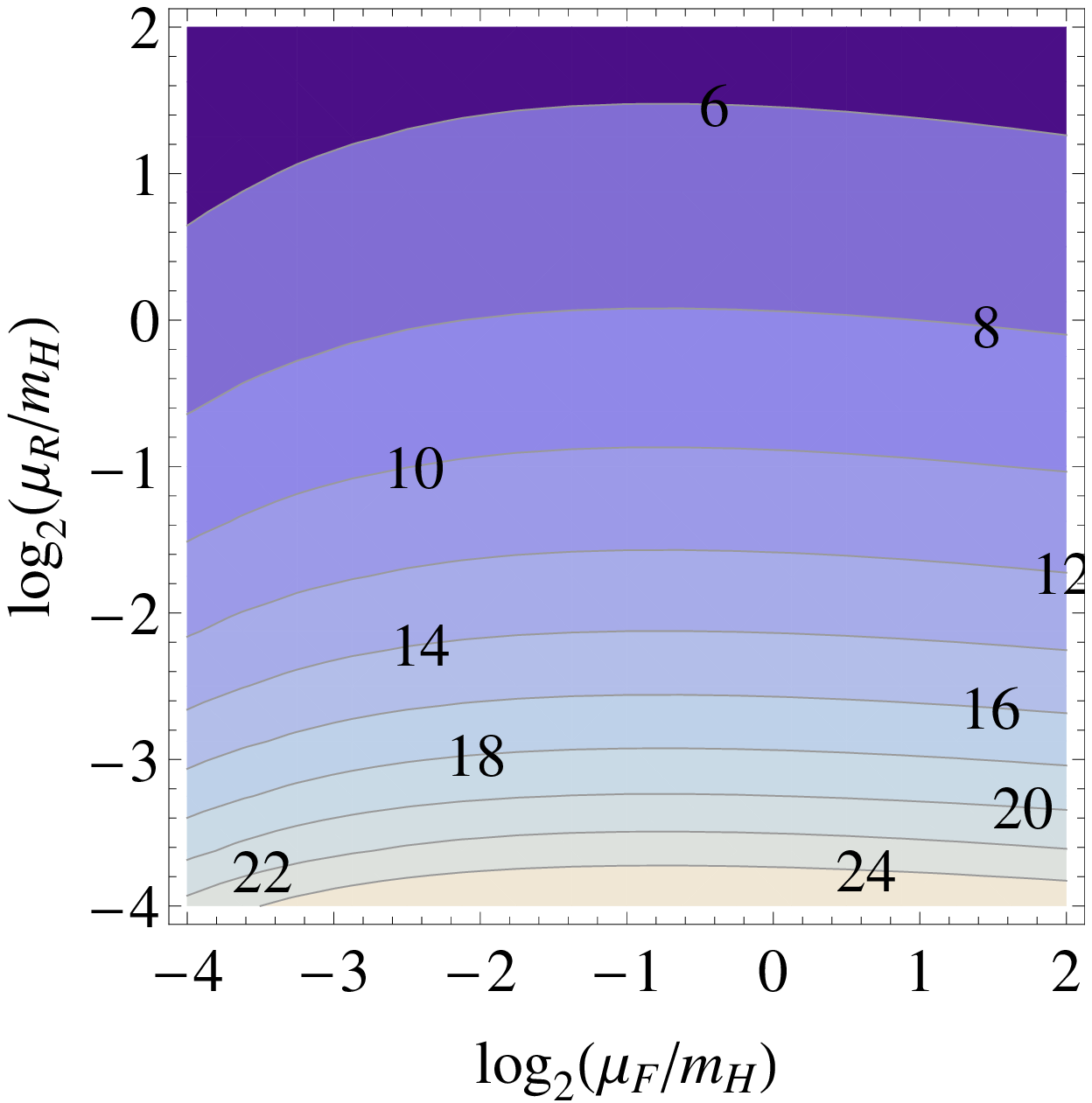}
    \caption{LO}
  \end{subfigure}%
  ~ 
  \begin{subfigure}[b]{0.30\textwidth}
    \centering
    \includegraphics[width=\textwidth]{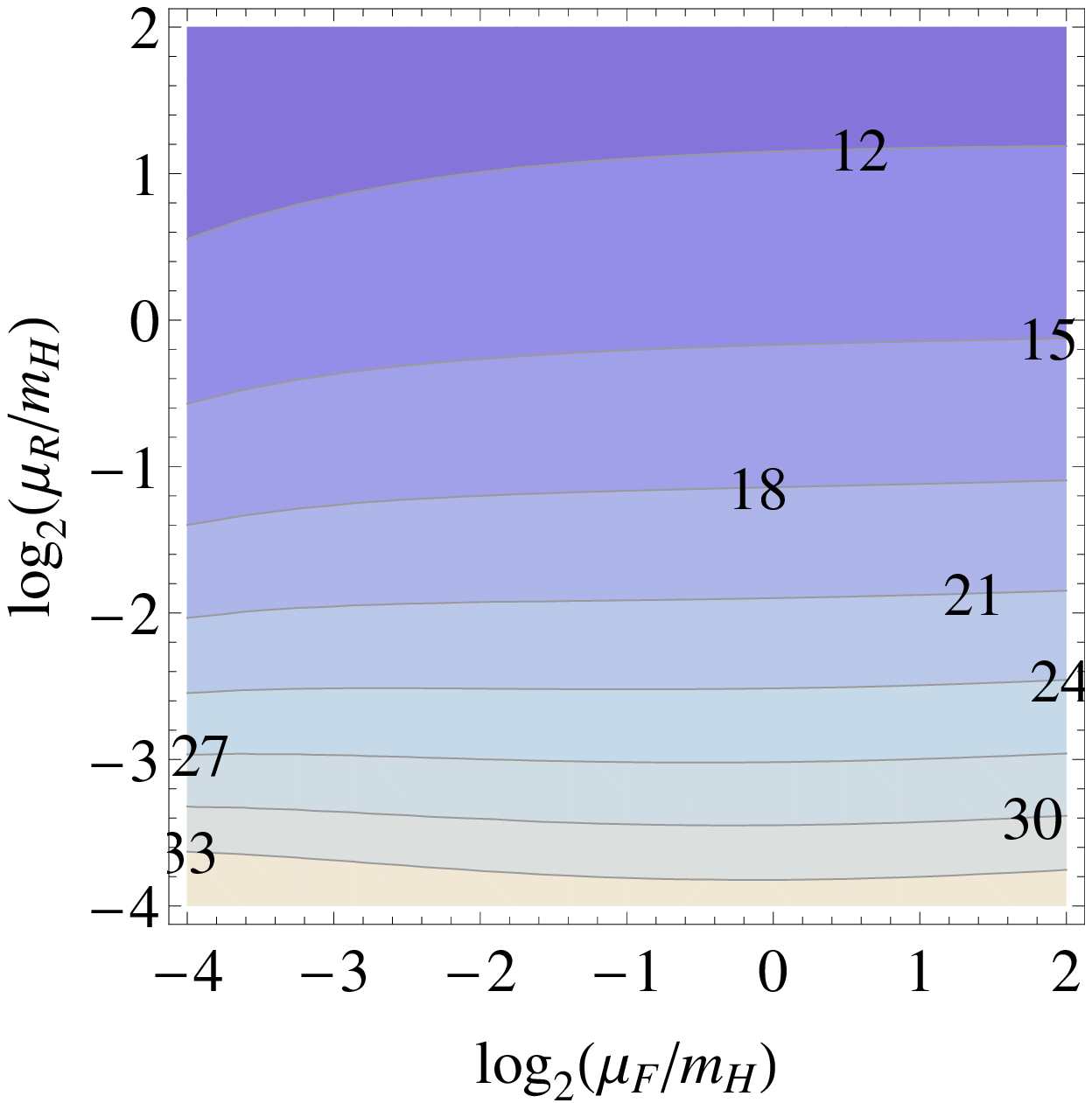}
    \caption{NLO}
  \end{subfigure}
  ~ 
  \begin{subfigure}[b]{0.30\textwidth}
    \centering
    \includegraphics[width=\textwidth]{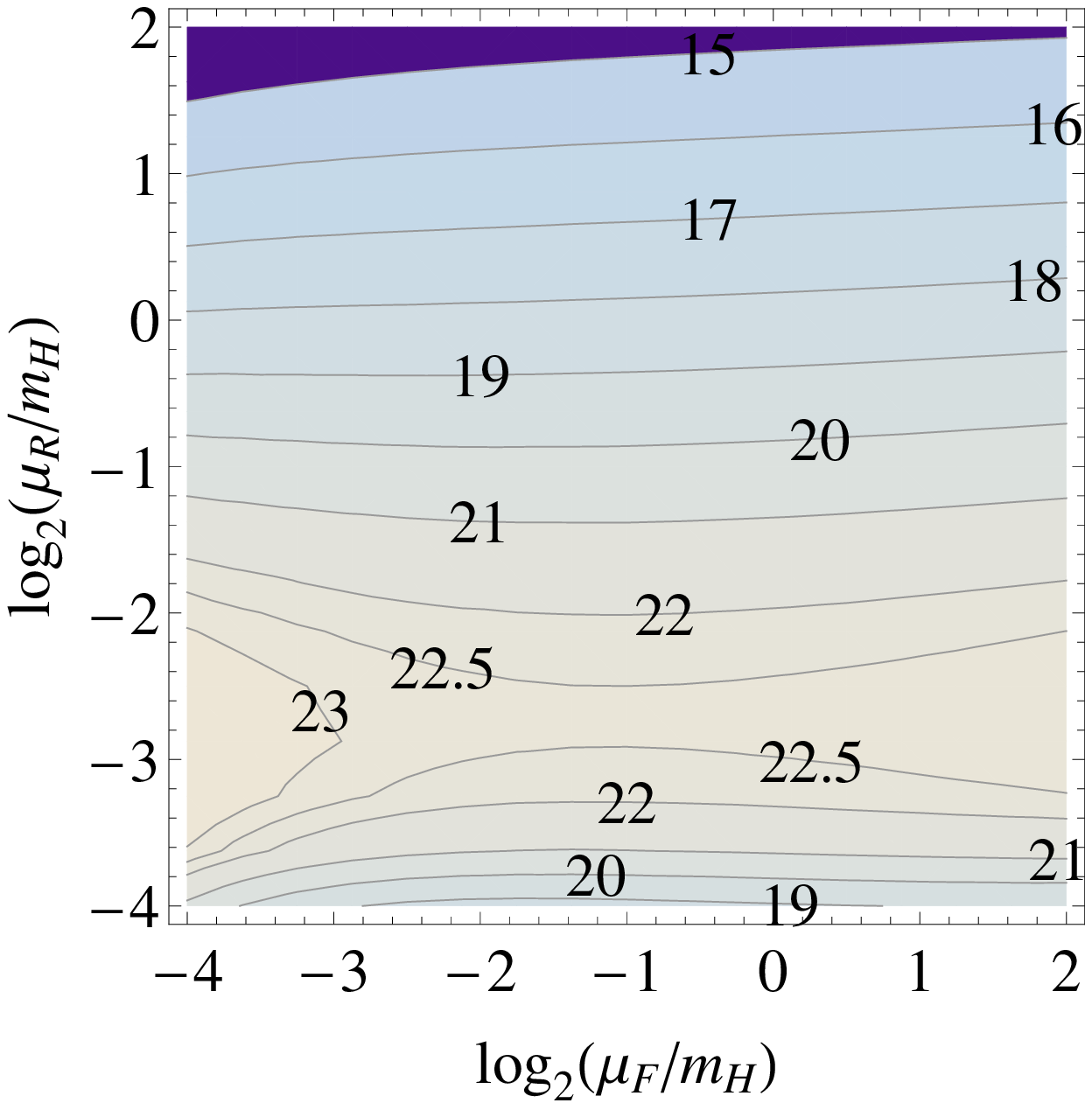}
    \caption{NNLO}
  \end{subfigure}
  \caption{2-D contour plots of the LO, NLO and NNLO cross section at the 8 TeV LHC. The value on the contours is the cross section in picobarns. The $x$-axis is $\log_2(\mu_f/m_h)$, the $y$-axis $\log_2(\mu_r/m_h)$. Our preferred central scale choice is located at (-1,-1).}
  \label{contourplots1}
\end{figure}

\begin{figure}
  \centering
  \begin{subfigure}[b]{0.30\textwidth}
    \centering
    \includegraphics[width=\textwidth]{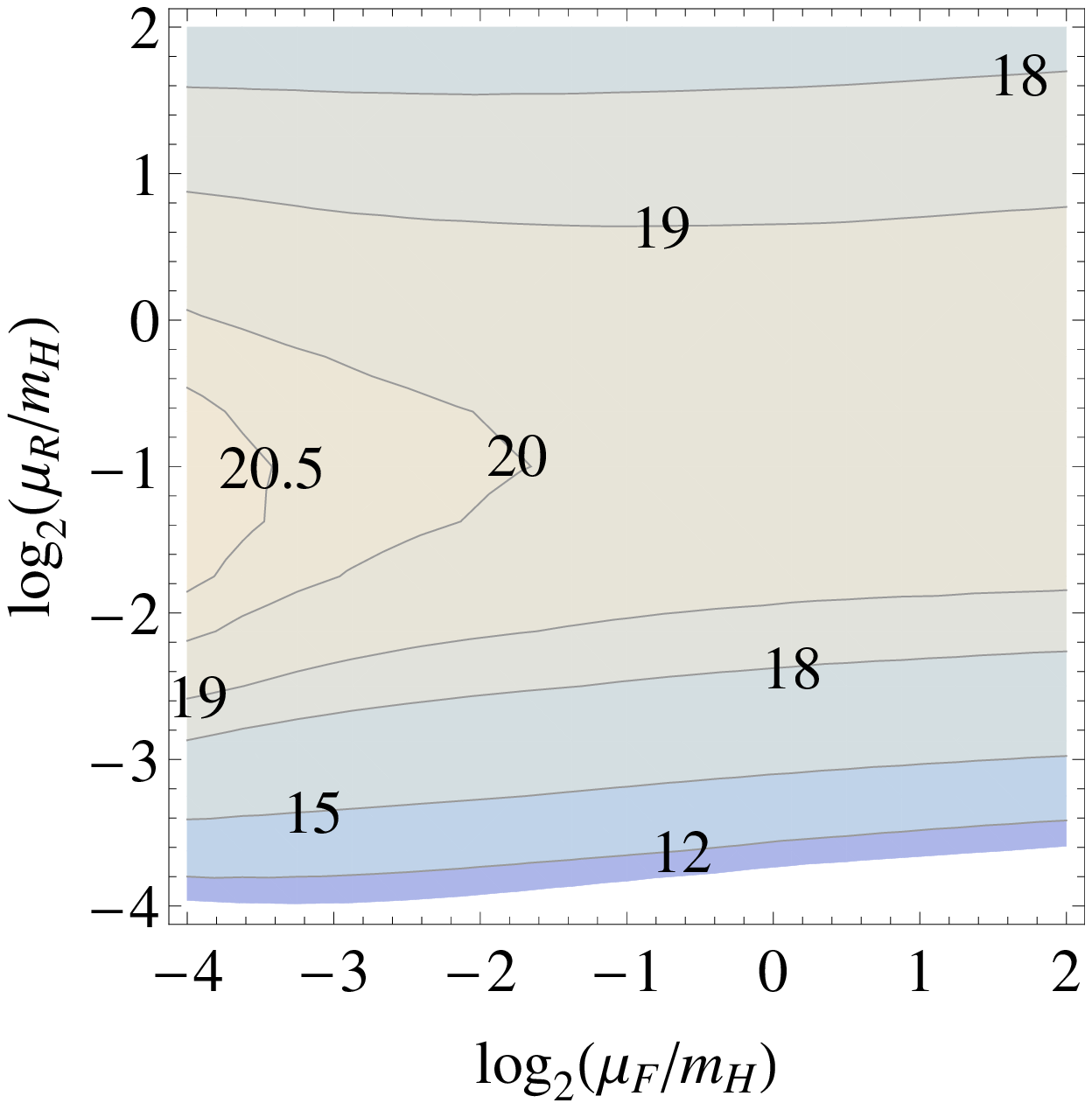}
    \caption{\n3lo ($K=10$)}
  \end{subfigure}%
  ~ 
  \begin{subfigure}[b]{0.30\textwidth}
    \centering
    \includegraphics[width=\textwidth]{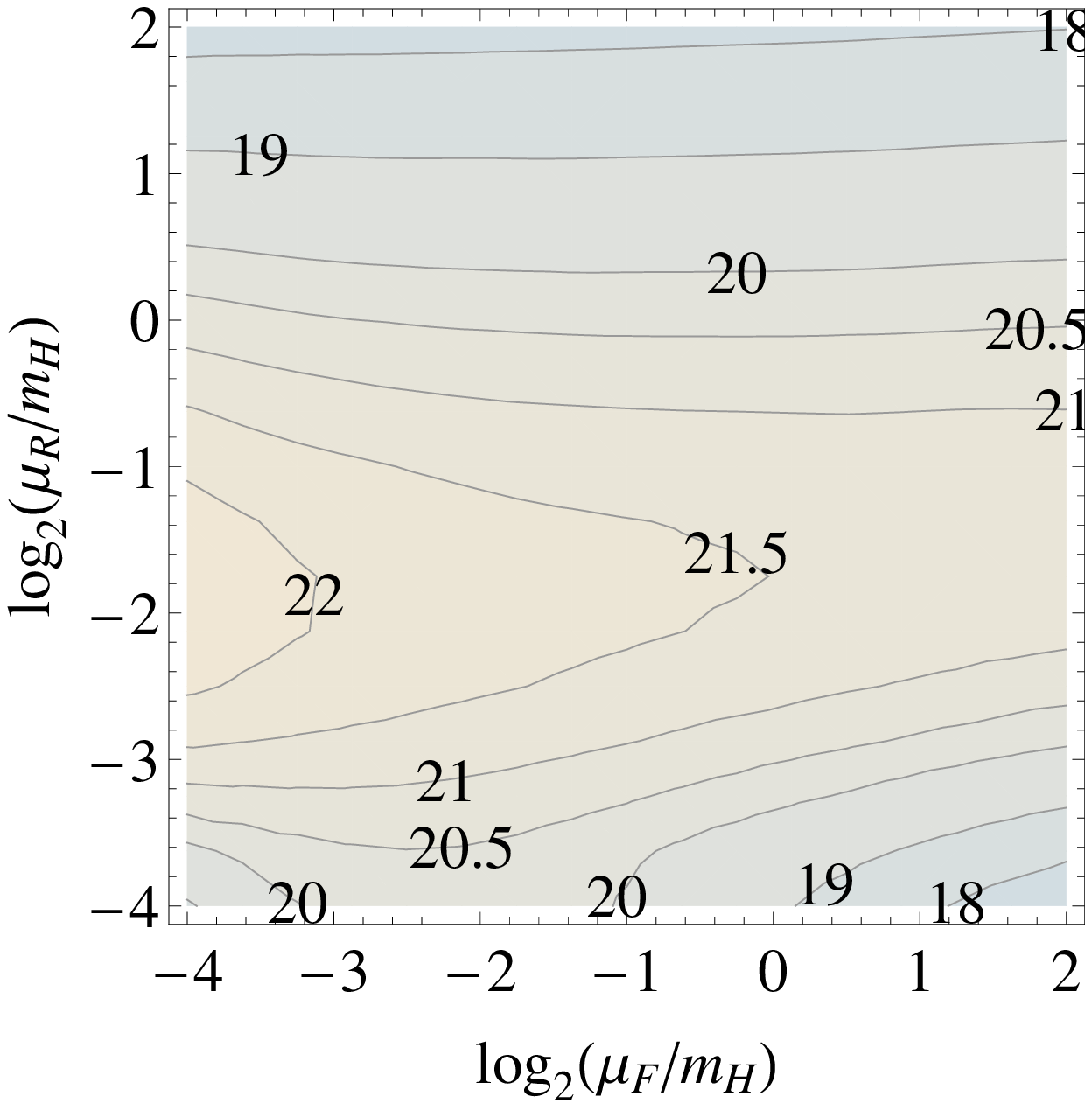}
    \caption{\n3lo ($K=20$)}
  \end{subfigure}
  ~ 
  \begin{subfigure}[b]{0.30\textwidth}
    \centering
    \includegraphics[width=\textwidth]{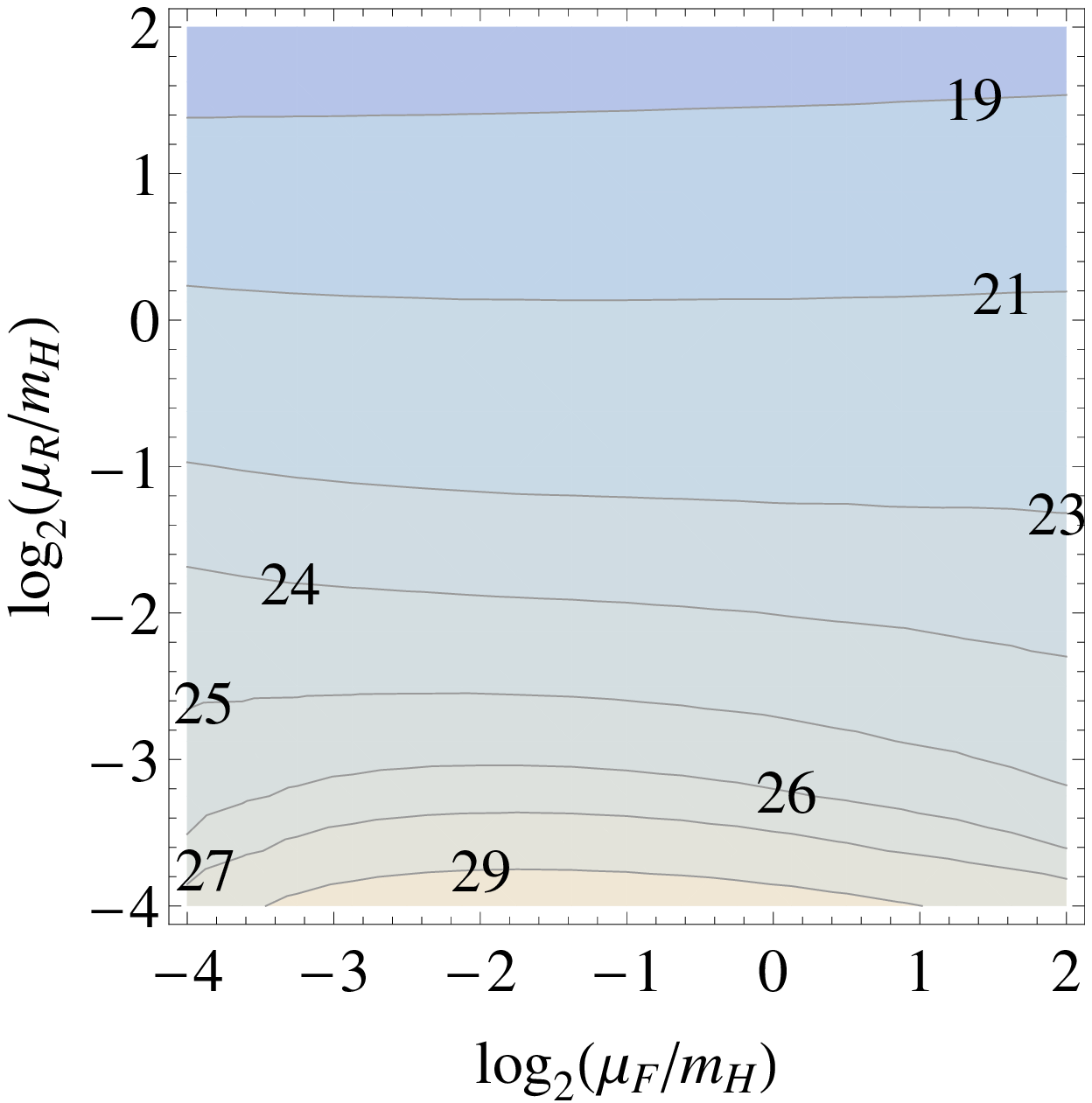}
    \caption{\n3lo ($K=30$)}
  \end{subfigure}
  \caption{2-D contour plots of the approximated \n3lo cross section at the 8 TeV LHC. The value on the contours is the cross section in picobarns. The $x$-axis is $\log_2(\mu_f/m_h)$, the $y$-axis $\log_2(\mu_r/m_h)$. Our preferred central scale choice is located at (-1,-1).}
  \label{contourplots2}
\end{figure}

The choice of the central scale around which the variation is performed has been an issue of debate lately, since different choices result in slightly different scale uncertainty estimates but also in different central values for the cross section. The choice is largely arbitrary, but various indications (like improved perturbative convergence, typical transverse momentum scales for radiated gluons, average Higgs transverse momentum etc.) point to a central scale choice that is lower than the traditional one at $m_h$, closer to $m_h / 2$. An alternative indication comes from the considerations of~\cite{Olness:2009qd}, where it is argued, looking at examples from jet physics, that a reasonable indication would be the position of the saddle point in a contour plot of the cross section as a function of $\mu_r$ and $\mu_f$. In figs.~\ref{contourplots1} and \ref{contourplots2} we show such contour plots for Higgs production at LO, NLO, NNLO and \n3lo (for three values of the parameter $K$). In the cases where a saddle point exists, its position points indeed to lower scale choices, and in the cases without a saddle point the plateau region is also located in lower scales. Given the extremely mild factorisation scale dependence, the saddle point or plateau region is largely determined by the $\mu_r$ plateau in all previous figures. 
%
%
%
%
%%%%%%%%%%%%%%%%%%%%%%%%%%%%%%%%%%%%%%%%%%%%%%%%%
%%%%% Conclusion
%%%%%%%%%%%%%%%%%%%%%%%%%%%%%%%%%%%%%%%%%%%%%%%%%%
\section{Conclusions}
\label{sec:Conclusions}
In this work we have presented all convolutions of lower-order partonic cross sections
and splitting kernels that contribute to order $a^5$ to Higgs production in gluon fusion. The results agree with the ones previously published in~\cite{Hoschele:2012xc}. Apart from the full expressions, we also provide all convolutions
expanded around threshold, as the full \n3lo corrections in this limit seem to be feasible in the near future.

We have also anticipated the scale dependence of the \n3lo gluon fusion cross section, into which the calculated convolutions enter. As is the case at NNLO, the factorisation scale dependence is extremely mild, at the per mille level or below. The overall scale uncertainty is driven by the renormalisation scale variation. The definite uncertainties depend on the size of the missing pure \n3lo contributions.  Scanning over a reasonable range for these contributions, we find that in the residual scale uncertainty can vary from $~2\% -  8\%$ depending on the magnitude of the hard real corrections, whose computation is, to our view, a prerequisite for a solid estimate of the \n3lo scale uncertainty. 

\section*{Acknowledgements}
We are very grateful to Babis Anastasiou for many fruitful discussions, as well as providing {\sc Maple} code for the partonic cross sections through NNLO.
Many thanks go to Claude Duhr for sharing his {\sc Mathematica} libraries and assisting in their usage, and to Davison Soper and Stephen Ellis for stimulating discussions on scale uncertainties. SB would like to thank Franz Herzog for pointing out the way of obtaining plus-plus convolutions by expanding a hypergeometric function.\\
This work was supported by the Swiss National Foundation under contract SNF 200020-126632.%Plots were created either with \verb+Mathematica+ or the root package.
 \vskip.5cm

%%%%%%%%%%%%%%%%%%%%%%%%%%%%%%%%%%%%%%%%%%%%%%%%%
%%%%% Appendices
%%%%%%%%%%%%%%%%%%%%%%%%%%%%%%%%%%%%%%%%%%%%%%%%%%
\appendix
\section{Convolutions of two plus-distributions}
\label{app:plusplusconv}
In the convolutions needed for collinear counterterms  we face the problem of convolutions involving one or
 more plus-distributions. Here, we demonstrate how to obtain all convolutions containing two plus-distributions.
\beq
\left( \DD_n(1-x) \otimes \DD_m(1-y)\right) (z) = \int_0^1 dx \, dy \left[\frac{\log(1-x)^n}{1-x}\right]_+ \left[\frac{\log(1-y)^m}{1-y}\right]_+ \delta(xy-z)
\eeq
To find these convolutions for all values of $m$ and $n$, we consider the following convolution integral:
\beq
I_{ab}(z) :=\int_0^1 dx\,dy (1-x)^{-1+a\epsilon}(1-y)^{-1+b\epsilon} \delta(xy-z) \, .
\eeq
We use the delta function to get rid of $x$ and then remap the integral onto the unit interval:
\begin{align}
 I_{ab}(z) & = \int_z^1 dy\, \frac{1}{y} \left( 1-\frac{z}{y} \right)^{-1+a\epsilon} (1-y)^{-1+b\epsilon} = \int_z^1 dy \, y^{-a\epsilon}(y-z)^{-1+a\epsilon}(1-y)^{-1+b\epsilon}\nonumber\\
 & = \int_0^1 d\lambda \, (1-z)[z+(1-z)\lambda]^{-a\epsilon}[\lambda(1-z)]^{-1+a\epsilon}[(1-z)(1-\lambda)]^{-1+b\epsilon}\nonumber\\
 & = (1-z)^{-1+(a+b)\epsilon} \int_0^1 d\lambda \, [z+(1-z)\lambda]^{-a\epsilon} \lambda^{-1+a\epsilon}(1-\lambda)^{-1+b\epsilon}\nonumber\\
 & = (1-z)^{-1+(a+b)\epsilon} \int_0^1 d\lambda \, [(1-\lambda)+z\lambda]^{-a\epsilon} \lambda^{-1+b\epsilon}(1-\lambda)^{-1+a\epsilon}\nonumber\\
 \label{lhs}
 & = (1-z)^{-1+(a+b)\epsilon}\,B(a\epsilon,b\epsilon) \, {}_2F_1(a\epsilon,b\epsilon,(a+b)\epsilon;1-z) \, .
\end{align}
In the second to last step, we mapped $\lambda\mapsto 1-\lambda$ and in the last step, the Euler definition of the hypergeometric function was used,
\beq
B(b,c-b)\,{}_2F_1(a,b,c;z) = \int_0^1 dx \, x^{b-1} (1-x)^{c-b-1} \underbrace{(1-zx)}_{(1-x)+(1-z)x}{}^{-a} \, ,
\eeq
where $B(x,y)$ denotes the Euler Beta-function. On the other hand, we may also directly expand the integrands in $I_{ab}$ in terms of a delta function and a tower of plus-distributions,
\begin{align}
 I_{ab}(z) & = \int_0^1 dx\,dy \left(\frac{\delta(1-x)}{a\epsilon} + \sum_{n\geq 0}\frac{(a\epsilon)^n}{n!}\DD_n(1-x)\right) \times\nonumber\\
 & \times \left(\frac{\delta(1-y)}{b\epsilon} + \sum_{m\geq 0}\frac{(b\epsilon)^m}{m!}\DD_m(1-y)\right) \delta(xy-z)\nonumber\\
 & = \frac{\delta(1-z)}{ab\epsilon^2} + \frac{1}{a\epsilon}\sum_{n\geq 0} \frac{(b\epsilon)^n}{n!}\DD_n(1-z) \frac{1}{b\epsilon}\sum_{n\geq 0} \frac{(a\epsilon)^n}{n!}\DD_n(1-z) +\nonumber\\
 & + \sum_{n,m\geq 0} \frac{(a\epsilon)^n\,(b\epsilon)^m}{n!\,m!} \left( \DD_n(1-x) \otimes \DD_m(1-y)\right) (z) \, .
\end{align}
When we now expand the first term of eq \ref{lhs} in the same way,
\beq
(1-z)^{-1+(a+b)\epsilon} = \frac{\delta(1-z)}{(a+b)\epsilon} + \sum_{n\geq 0}\frac{((a+b)\epsilon)^n}{n!}\DD_n(1-z) \, ,
\eeq
and expand the Beta-function and the ${}_2F_1$ (using the {\sc Mathematica} Package \verb+HypExp+~\cite{Huber:2005yg}) in $\epsilon$ as well, we can equate the two sides order by order in $\epsilon$.\\
The double and single poles cancel and for $\mathcal{O}(\epsilon^0)$ we find the equation
\beq
\left( \DD_0(1-x) \otimes \DD_0(1-y)\right) (z) = -\frac{\pi^2}{6}\delta(1-z) + 2\DD_1(1-z)-\frac{\log(z)}{1-z} \, .
\eeq
 Higher orders in $\epsilon$ of the equation contain more than one plus-plus-convolutions, but
they can be isolated by extracting the corresponding coefficient of $a$ and $b$. To find the expression for the convolution
 $\left( \DD_n(1-x) \otimes \DD_m(1-y)\right) (z)$, one has to take the $\mathcal{O}(\epsilon^{n+m} a^n b^m)$ coefficient of the equation,
 or equivalently the $\mathcal{O}(\epsilon^{n+m} a^m b^n)$ coefficient since the expressions are symmetric in $a$ and $b$.

In this way, we found all the plus-plus convolutions needed for this work, which are listed here for completeness.
\begin{align}
\left( \DD_0 \otimes \DD_0\right) (z)  = & -\frac{\pi^2}{6}\delta(1-z) + 2\DD_1(1-z)-\frac{\log(z)}{1-z}
\end{align}
\begin{align}
\left( \DD_0 \otimes \DD_1\right) (z)  = & \zeta_3\,\delta(1-z) - \frac{\pi^2}{6}\DD_0(1-z) + \frac{3}{2}\DD_2(1-z) -\frac{\log(z)\log(1-z)}{1-z} 
\end{align}
\begin{align}
\left( \DD_0 \otimes \DD_2\right) (z)  = & \frac{\pi^4}{45}\delta(1-z) + \frac{4}{3}\DD_3(1-z) - \frac{\pi^2}{3}\DD_1(1-z) + 2\zeta_3\,\DD_0(1-z)\nonumber \\
 &  -\frac{\log(z)\log^2(1-z)}{1-z} 
\end{align}
\begin{align}
\left( \DD_1 \otimes \DD_1\right) (z)  = & -\frac{\pi^4}{360}\delta(1-z) + \DD_3(1-z) - \frac{\pi^2}{3}\DD_1(1-z)  + 2\zeta_3\,\DD_0(1-z) \nonumber \\
 & -\frac{\log(z)\log^2(1-z)}{1-z} -\frac{\log(z)\Li_2(z)}{1-z} + 2\frac{\Li_3(z)-\zeta_3}{1-z}
\end{align}
\begin{align}
\left( \DD_0 \otimes \DD_3\right) (z)  = & 6\zeta_5\,\delta(1-z) + \frac{5}{4}\DD_4(1-z) - \frac{\pi^2}{2}\DD_2(1-z) \nonumber \\
 & + 6\zeta_3\,\DD_1(1-z) -\frac{\pi^4}{15}\DD_0(1-z) -\frac{\log(z)\log^3(1-z)}{1-z} 
\end{align}
\begin{align}
\left( \DD_1 \otimes \DD_2\right) (z)  = & \left(4\zeta_5 -\frac{\pi^2\zeta_3}{3} \right)\delta(1-z) + \frac{5}{6}\DD_4(1-z) - \frac{\pi^2}{2}\DD_2(1-z) \nonumber \\
 & + 6\zeta_3\,\DD_1(1-z) -\frac{\pi^4}{36}\DD_0(1-z) + \frac{\Li_2^2(1-z)}{1-z} \nonumber \\
 & +  4\frac{\log(1-z)(\Li_3(z)-\zeta_3)}{1-z} + \frac{\log(z)}{1-z}\Big[2\log(z)\log^2(1-z) \nonumber \\
 & - \log^3(1-z) -\frac{\pi^2}{3}\log(1-z) + 2\log(1-z)\Li_2(1-z) \nonumber \\
 & + 2\Li_3(1-z) - 2\zeta_3 \Big]
\end{align}
\begin{align}
 \left( \DD_0 \otimes \DD_4\right) (z)  = & -\frac{8\pi^6}{315}\delta(1-z) + \frac{6}{5}\DD_5(1-z) - \frac{2\pi^2}{3}\DD_3(1-z) \nonumber \\
 & + 12\zeta_3\,\DD_2(1-z) -\frac{4\pi^4}{15}\DD_1(1-z) +24\zeta_5\,\DD_0(1-z) \nonumber \\
 & -\frac{\log(z)\log^4(1-z)}{1-z} 
\end{align}
\begin{align}
\left( \DD_1 \otimes \DD_3\right) (z)  = & \left(3\zeta_3^2 -\frac{\pi^6}{210} \right)\delta(1-z) + \frac{3}{4}\DD_5(1-z) - \frac{2\pi^2}{3}\DD_3(1-z) \nonumber \\
 & + 12\zeta_3\,\DD_2(1-z) -\frac{3\pi^4}{20}\DD_1(1-z) \left(18\zeta_5-\pi^2\zeta_3\right)\DD_0(1-z) \nonumber \\
 & + \frac{1}{1-z} \Bigg[ 3\log^3(1-z)\log^2(z) + \log(z) \Bigg(\frac{\pi^4}{15} - \frac{\pi^2\log^2(1-z)}{2} \nonumber \\
 & - \log^4(1-z) + 3\log^2(1-z)\Li_2(1-z) + 6\log(1-z)\Li_3(1-z)  \nonumber \\ 
 & - 6\Li_4(1-z) - 6\zeta_3\log(1-z) \Bigg) + 12\mathrm{H}(0,0,1,0,1;1-z)   \nonumber \\ 
 & + 24\mathrm{H}(0,0,0,1,1;1-z) + 6\log^2(1-z)(\Li_3(z)-\zeta_3)   \nonumber \\ 
 & + 3\log(1-z)\Li_2^2(1-z) - 6\Li_2(1-z)\Li_3(1-z) \Bigg]
\end{align}
The above expressions agree with the ones given in \cite{Hoschele:2012xc} (eq.~22) and \cite{Ridder:2012dg} (eq.~C.28 - C.31). For the cases $\DD_0\otimes\DD_n$, the combination of 
harmonic polylogarithms given in the references collapses to the single term $-\log(z)\log^n(1-z)/(1-z)$.
\section{Ancillary files}
\label{app:files}
Here we briefly describe the files accompanying the publication. All files are available both in {\sc Maple} (\emph{file.mpl}) and {\sc Mathematica} (\emph{file.m}) format. 
\begin{enumerate}
\item \verb+sigma.m+ and \verb+sigma.mpl+: Contain the partonic cross sections $\tilde{\sigma}^{(n)}_{ij}$ through the respective order in $\epsilon$ needed for the \n3lo cross
section. There are five different partonic channels ($gg$, $qg$, $q\bar{q}$, $qq$ and $qQ$, where $q \neq Q\neq \bar{q}$). Only the $gg$ channel contains terms of soft origin
($\delta(1-x)$ and $\DD_n(1-x)$ terms). They are denoted by \verb+d1(1-x)+ and \verb+DD(n,x)+, respectively.\\
Harmonic polylogarithms are denoted by $H$, and can be cast in the form used by the package HPL~\cite{Maitre:2005uu,Maitre:2007kp} via the {\sc Mathematica} replacement rule\\
\verb+H[a__,x_] -> HPL[{a},x]+.
\item \verb+convolutions.m+ and \verb+convolutions.mpl+: Contain the 80 convolutions of splitting kernels and partonic cross sections required for the \n3lo cross section, to
the respective order in $\epsilon$ needed. The names of the convolutions are simply the concatenation of all ingredients. The convolution $P^{(0)}_{gg}\otimes P^{(0)}_{gq}\otimes \tilde{\sigma}^{(0)}_{gg}$
for example is called \verb+pgg0pgq0sigma0gg+.\\
All expressions are given in terms of soft terms like $\delta(1-x)$ and $\DD_n(1-x)$, where present, and HPLs for the regular parts.
\item \verb+convolutions_softlimit.m+ and \verb+convolutions_softlimit.mpl+: Contain the same 80 convolutions, but the regular parts are expanded in the variable $xp\equiv 1-x$ as described in section
\ref{sec:softlimits}. The names are prepended by a capital \verb+S+ in this file, e.g. \verb+Spgg0pgq0sigma0gg+ for the soft limit of the convolution $P^{(0)}_{gg}\otimes P^{(0)}_{gq}\otimes \tilde{\sigma}^{(0)}_{gg}$.
\item \verb+splittingkernels.m+ and \verb+splittingkernels.mpl+: Contain the twelve spltting kernels listed in section~\ref{sec:ingredients}, in the conventions given in
section~\ref{sec:AP}. Furthermore, the intermediate double and triple convolutions among one-loop and two-loop kernels are provided, as well.
All expressions are given in terms of soft terms like $\delta(1-x)$ and $\DD_n(1-x)$, where present, and HPLs for the regular parts.
\end{enumerate}
%
%%%%%%%%%%%%%%%%%%%%%%%%%%%%%%%%%%%%%%%%%%%%%%%%%
%%%%% bibliography
%%%%%%%%%%%%%%%%%%%%%%%%%%%%%%%%%%%%%%%%%%%%%%%%%%

%\pagebreak
\bibliographystyle{JHEP}

\providecommand{\href}[2]{#2}\begingroup\raggedright
\endgroup

\end{document}

%% file: data/table_mstw.tex
\begin{tabular}{lccc}
\toprule
Order & Cross section [pb] & $\sigma/\sigma_{\text{NNLO}}$ & $\sigma/\sigma_{\text{LO}}$ \\ 
\midrule
LO & $10.31\;{}^{+26.9\%}_{-16.6\%}$ & $0.51$ & $1.00$ \\ \addlinespace
NLO & $17.41\;{}^{+20.8\%}_{-12.7\%}$ & $0.86$ & $1.69$ \\ \addlinespace
NNLO & $20.27\;{}^{+8.3\%}_{-7.1\%}$ & $1.00$ & $1.97$ \\ \addlinespace
\n3lo (K=0) & $18.53\;{}^{+1.2\%}_{-7.9\%}$ & $0.91$ & $1.80$ \\ \addlinespace
\n3lo (K=5) & $19.23\;{}^{+0.3\%}_{-5.1\%}$ & $0.95$ & $1.87$ \\ \addlinespace
\n3lo (K=10) & $19.92\;{}^{+0.0\%}_{-2.6\%}$ & $0.98$ & $1.93$ \\ \addlinespace
\n3lo (K=15) & $20.62\;{}^{+0.4\%}_{-2.2\%}$ & $1.02$ & $2.00$ \\ \addlinespace
\n3lo (K=20) & $21.31\;{}^{+2.0\%}_{-3.1\%}$ & $1.05$ & $2.07$ \\ \addlinespace
\n3lo (K=30) & $22.70\;{}^{+6.0\%}_{-4.9\%}$ & $1.12$ & $2.20$ \\ \addlinespace
\n3lo (K=40) & $24.09\;{}^{+9.6\%}_{-6.5\%}$ & $1.19$ & $2.34$ \\ \addlinespace
\bottomrule
\end{tabular}

%% file: paper.bbl
\begin{thebibliography}{100}

%\cite{Chatrchyan:2012ufa}
\bibitem{Chatrchyan:2012ufa}
  S.~Chatrchyan {\it et al.}  [CMS Collaboration],
  %``Observation of a new boson at a mass of 125 GeV with the CMS experiment at the LHC,''
  Phys.\ Lett.\ B {\bf 716} (2012) 30
  [arXiv:1207.7235 [hep-ex]].
  %%CITATION = ARXIV:1207.7235;%%

%\cite{Aad:2012tfa}
\bibitem{Aad:2012tfa}
  G.~Aad {\it et al.}  [ATLAS Collaboration],
  %``Observation of a new particle in the search for the Standard Model Higgs boson with the ATLAS detector at the LHC,''
  Phys.\ Lett.\ B {\bf 716} (2012) 1
  [arXiv:1207.7214 [hep-ex]].
  %%CITATION = ARXIV:1207.7214;%%
  %728 citations counted in INSPIRE as of 28 Feb 2013

%\cite{Dawson:1990zj}
\bibitem{Dawson:1990zj}
  S.~Dawson,
  %``Radiative corrections to Higgs boson production,''
  Nucl.\ Phys.\  B {\bf 359}, 283 (1991).
  %%CITATION = NUPHA,B359,283;%%

%\cite{Djouadi:1991tka}
\bibitem{Djouadi:1991tka}
  A.~Djouadi, M.~Spira and P.~M.~Zerwas,
  %``Production of Higgs bosons in proton colliders: QCD corrections,''
  Phys.\ Lett.\  B {\bf 264} (1991) 440.
  %%CITATION = PHLTA,B264,440;%%

%\cite{Graudenz:1992pv}
\bibitem{Graudenz:1992pv}
  D.~Graudenz, M.~Spira and P.~M.~Zerwas,
  %``QCD corrections to Higgs boson production at proton proton colliders,''
  Phys.\ Rev.\ Lett.\  {\bf 70} (1993) 1372.
  %%CITATION = PRLTA,70,1372;%%

%\cite{Spira:1995rr}
\bibitem{Spira:1995rr}
  M.~Spira, A.~Djouadi, D.~Graudenz and P.~M.~Zerwas,
  %``Higgs boson production at the LHC,''
  Nucl.\ Phys.\  B {\bf 453}, 17 (1995)
  [arXiv:hep-ph/9504378].
  %%CITATION = NUPHA,B453,17;%%

%cite{Anastasiou:2006hc}
\bibitem{Anastasiou:2006hc}
  C.~Anastasiou, S.~Beerli, S.~Bucherer, A.~Daleo and Z.~Kunszt,
  %``Two-loop amplitudes and master integrals for the production of a Higgs
  %boson via a massive quark and a scalar-quark loop,''
  JHEP {\bf 0701} (2007) 082
  [arXiv:hep-ph/0611236].
  %%CITATION = JHEPA,0701,082;%%

%\cite{Aglietti:2006tp}
\bibitem{Aglietti:2006tp}
  U.~Aglietti, R.~Bonciani, G.~Degrassi and A.~Vicini,
  %``Analytic Results for Virtual QCD Corrections to Higgs Production and Decay,''
  JHEP {\bf 0701}, 021 (2007)
  [arXiv:hep-ph/0611266].
  %%CITATION = JHEPA,0701,021;%%

%\cite{Bonciani:2007ex}
\bibitem{Bonciani:2007ex}
 R.~Bonciani, G.~Degrassi and A.~Vicini,
 %``Scalar particle contribution to Higgs production via gluon fusion at NLO,''
 JHEP {\bf 0711} (2007) 095
 [arXiv:0709.4227 [hep-ph]].
 %%CITATION = ARXIV:0709.4227;%%

%\cite{Harlander:2002wh}
\bibitem{Harlander:2002wh} 
  R.~V.~Harlander and W.~B.~Kilgore,
  %``Next-to-next-to-leading order Higgs production at hadron colliders,''
  Phys.\ Rev.\ Lett.\  {\bf 88}, 201801 (2002)
  [hep-ph/0201206].
  %%CITATION = HEP-PH/0201206;%%

%\cite{Anastasiou:2002yz}
\bibitem{Anastasiou:2002yz}
  C.~Anastasiou and K.~Melnikov,
  %``Higgs boson production at hadron colliders in NNLO QCD,''
  Nucl.\ Phys.\ B {\bf 646} (2002) 220
  [hep-ph/0207004].
  %%CITATION = HEP-PH/0207004;%%

%\cite{Ravindran:2003um}
\bibitem{Ravindran:2003um} 
  V.~Ravindran, J.~Smith and W.~L.~van Neerven,
  %``NNLO corrections to the total cross-section for Higgs boson production in hadron hadron collisions,''
  Nucl.\ Phys.\ B {\bf 665}, 325 (2003)
  [hep-ph/0302135].
  %%CITATION = HEP-PH/0302135;%%

%\cite{Aglietti:2004nj}
\bibitem{Aglietti:2004nj}
 U.~Aglietti, R.~Bonciani, G.~Degrassi and A.~Vicini,
 %``Two loop light fermion contribution to Higgs production and decays,''
 Phys.\ Lett.\ B {\bf 595} (2004) 432
 [hep-ph/0404071].
 %%CITATION = HEP-PH/0404071;%%

%\cite{Degrassi:2004mx}
\bibitem{Degrassi:2004mx}
 G.~Degrassi and F.~Maltoni,
 %``Two-loop electroweak corrections to Higgs production at hadron colliders,''
 Phys.\ Lett.\ B {\bf 600} (2004) 255
 [hep-ph/0407249].
 %%CITATION = HEP-PH/0407249;%%
 %98 citations counted in INSPIRE as of 11 Jun 2013

%\cite{Actis:2008ug,Actis:2008ts}
\bibitem{Actis:2008ug}
  S.~Actis, G.~Passarino, C.~Sturm and S.~Uccirati,
  %``NLO Electroweak Corrections to Higgs Boson Production at Hadron Colliders,''
  Phys.\ Lett.\ B {\bf 670} (2008) 12
  [arXiv:0809.1301 [hep-ph]].
  %%CITATION = ARXIV:0809.1301;%%
  %196 citations counted in INSPIRE as of 08 May 2013

%\cite{Actis:2008ts}
\bibitem{Actis:2008ts}
  S.~Actis, G.~Passarino, C.~Sturm and S.~Uccirati,
  %``NNLO Computational Techniques: The Cases H ---> gamma gamma and H ---> g g,''
  Nucl.\ Phys.\ B {\bf 811} (2009) 182
  [arXiv:0809.3667 [hep-ph]].
  %%CITATION = ARXIV:0809.3667;%%
  %90 citations counted in INSPIRE as of 08 May 2013

%\cite{Anastasiou:2008tj}
\bibitem{Anastasiou:2008tj}
  C.~Anastasiou, R.~Boughezal and F.~Petriello,
  %``Mixed QCD-electroweak corrections to Higgs boson production in gluon fusion,''
  JHEP {\bf 0904} (2009) 003
  [arXiv:0811.3458 [hep-ph]].
  %%CITATION = ARXIV:0811.3458;%%
  %235 citations counted in INSPIRE as of 08 May 2013

%\cite{Brein:2010xj}
\bibitem{Brein:2010xj}
  O.~Brein,
  %``Electroweak and Bottom Quark Contributions to Higgs Boson plus Jet Production,''
  Phys.\ Rev.\ D {\bf 81} (2010) 093006
  [arXiv:1003.4438 [hep-ph]].
  %%CITATION = ARXIV:1003.4438;%%
  %13 citations counted in INSPIRE as of 08 May 2013

%\cite{Boughezal:2013uia}
\bibitem{Boughezal:2013uia}
  R.~Boughezal, F.~Caola, K.~Melnikov, F.~Petriello and M.~Schulze,
  %``Higgs boson production in association with a jet at next-to-next-to-leading order in perturbative QCD,''
  arXiv:1302.6216 [hep-ph].
  %%CITATION = ARXIV:1302.6216;%%
  %4 citations counted in INSPIRE as of 27 May 2013
  
%\cite{Anastasiou:2011pi}
\bibitem{Anastasiou:2011pi}
  C.~Anastasiou, S.~Buehler, F.~Herzog and A.~Lazopoulos,
  %``Total cross-section for Higgs boson hadroproduction with anomalous Standard Model interactions,''
  JHEP {\bf 1112} (2011) 058
  [arXiv:1107.0683 [hep-ph]].
  %%CITATION = ARXIV:1107.0683;%%

%\cite{Marzani:2008az}
\bibitem{Marzani:2008az}
 S.~Marzani, R.~D.~Ball, V.~Del Duca, S.~Forte and A.~Vicini,
 %``Higgs production via gluon-gluon fusion with finite top mass beyond next-to-leading order,''
 Nucl.\ Phys.\ B {\bf 800} (2008) 127
 [arXiv:0801.2544 [hep-ph]].
 %%CITATION = ARXIV:0801.2544;%%

%\cite{Harlander:2009mq}
\bibitem{Harlander:2009mq} 
  R.~V.~Harlander and K.~J.~Ozeren,
  %``Finite top mass effects for hadronic Higgs production at next-to-next-to-leading order,''
  JHEP {\bf 0911}, 088 (2009)
  [arXiv:0909.3420 [hep-ph]].
  %%CITATION = ARXIV:0909.3420;%%
  
%\cite{Pak:2009dg}
\bibitem{Pak:2009dg} 
  A.~Pak, M.~Rogal and M.~Steinhauser,
  %``Finite top quark mass effects in NNLO Higgs boson production at LHC,''
  JHEP {\bf 1002}, 025 (2010)
  [arXiv:0911.4662 [hep-ph]].
  %%CITATION = ARXIV:0911.4662;%%


%\cite{Catani:2003zt}
\bibitem{Catani:2003zt}
  S.~Catani, D.~de Florian, M.~Grazzini and P.~Nason,
  %``Soft gluon resummation for Higgs boson production at hadron colliders,''
  JHEP {\bf 0307} (2003) 028
  [hep-ph/0306211].
  %%CITATION = HEP-PH/0306211;%%
  %383 citations counted in INSPIRE as of 08 May 2013

%\cite{deFlorian:2012yg}
\bibitem{deFlorian:2012yg}
  D.~de Florian and M.~Grazzini,
  %``Higgs production at the LHC: updated cross sections at $\sqrt{s}=8$ TeV,''
  Phys.\ Lett.\ B {\bf 718} (2012) 117
  [arXiv:1206.4133 [hep-ph]].
  %%CITATION = ARXIV:1206.4133;%%
  %25 citations counted in INSPIRE as of 14 May 2013


%\cite{Ahrens:2008qu,Ahrens:2008nc}
\bibitem{Ahrens:2008qu}
  V.~Ahrens, T.~Becher, M.~Neubert and L.~L.~Yang,
  %``Origin of the Large Perturbative Corrections to Higgs Production at Hadron Colliders,''
  Phys.\ Rev.\ D {\bf 79} (2009) 033013
  [arXiv:0808.3008 [hep-ph]].
  %%CITATION = ARXIV:0808.3008;%%
  %58 citations counted in INSPIRE as of 08 May 2013

%\cite{Ahrens:2008nc}
\bibitem{Ahrens:2008nc}
  V.~Ahrens, T.~Becher, M.~Neubert and L.~L.~Yang,
  %``Renormalization-Group Improved Prediction for Higgs Production at Hadron Colliders,''
  Eur.\ Phys.\ J.\ C {\bf 62} (2009) 333
  [arXiv:0809.4283 [hep-ph]].
  %%CITATION = ARXIV:0809.4283;%%
  %85 citations counted in INSPIRE as of 08 May 2013

%\cite{Ahrens:2010rs}
\bibitem{Ahrens:2010rs}
  V.~Ahrens, T.~Becher, M.~Neubert and L.~L.~Yang,
  %``Updated Predictions for Higgs Production at the Tevatron and the LHC,''
  Phys.\ Lett.\ B {\bf 698} (2011) 271
  [arXiv:1008.3162 [hep-ph]].
  %%CITATION = ARXIV:1008.3162;%%
  %18 citations counted in INSPIRE as of 14 May 2013
  
  %\cite{Peskin:2012we}
\bibitem{Peskin:2012we}
  M.~E.~Peskin,
  %``Comparison of LHC and ILC Capabilities for Higgs Boson Coupling Measurements,''
  arXiv:1207.2516 [hep-ph].
  %%CITATION = ARXIV:1207.2516;%%
  %35 citations counted in INSPIRE as of 13 May 2013
  
  %\cite{Klute:2012pu}
\bibitem{Klute:2012pu}
  M.~Klute, R.~Lafaye, T.~Plehn, M.~Rauch and D.~Zerwas,
  %``Measuring Higgs Couplings from LHC Data,''
  Phys.\ Rev.\ Lett.\  {\bf 109} (2012) 101801
  [arXiv:1205.2699 [hep-ph]].
  %%CITATION = ARXIV:1205.2699;%%
  %63 citations counted in INSPIRE as of 13 May 2013
  
  %\cite{Klute:2013cx}
\bibitem{Klute:2013cx}
  M.~Klute, R.~Lafaye, T.~Plehn, M.~Rauch and D.~Zerwas,
  %``Measuring Higgs Couplings at a Linear Collider,''
  Europhys.\ Lett.\  {\bf 101} (2013) 51001
  [arXiv:1301.1322 [hep-ph]].
  %%CITATION = ARXIV:1301.1322;%%
  %2 citations counted in INSPIRE as of 13 May 2013

%\cite{Baikov:2006ch}
\bibitem{Baikov:2006ch}
  P.~A.~Baikov and K.~G.~Chetyrkin,
  %``Higgs Decay into Hadrons to Order alpha**5(s),''
  Phys.\ Rev.\ Lett.\  {\bf 97} (2006) 061803
  [hep-ph/0604194].
  %%CITATION = HEP-PH/0604194;%%
  %25 citations counted in INSPIRE as of 08 May 2013

%\cite{Moch:2005tm}
\bibitem{Moch:2005tm}
  S.~Moch, J.~A.~M.~Vermaseren and A.~Vogt,
  %``Three-loop results for quark and gluon form-factors,''
  Phys.\ Lett.\ B {\bf 625} (2005) 245
  [hep-ph/0508055].
  %%CITATION = HEP-PH/0508055;%%

 %\cite{Baikov:2009bg}
\bibitem{Baikov:2009bg}
  P.~A.~Baikov, K.~G.~Chetyrkin, A.~V.~Smirnov, V.~A.~Smirnov and M.~Steinhauser,
  %``Quark and gluon form factors to three loops,''
  Phys.\ Rev.\ Lett.\  {\bf 102} (2009) 212002
  [arXiv:0902.3519 [hep-ph]].
  %%CITATION = ARXIV:0902.3519;%%

 %\cite{Lee:2010cga}
\bibitem{Lee:2010cga}
  R.~N.~Lee, A.~V.~Smirnov and V.~A.~Smirnov,
  %``Analytic Results for Massless Three-Loop Form Factors,''
  JHEP {\bf 1004} (2010) 020
  [arXiv:1001.2887 [hep-ph]].
  %% CITATION = ARXIV:1001.2887;%%
  
  %\cite{Gehrmann:2010ue}
\bibitem{Gehrmann:2010ue}
  T.~Gehrmann, E.~W.~N.~Glover, T.~Huber, N.~Ikizlerli and C.~Studerus,
  %``Calculation of the quark and gluon form factors to three loops in QCD,''
  JHEP {\bf 1006} (2010) 094
  [arXiv:1004.3653 [hep-ph]].
  %%CITATION = ARXIV:1004.3653;%%

%\cite{Moch:2005ky}
\bibitem{Moch:2005ky}
  S.~Moch and A.~Vogt,
  %``Higher-order soft corrections to lepton pair and Higgs boson production,''
  Phys.\ Lett.\ B {\bf 631} (2005) 48
  [hep-ph/0508265].
  %%CITATION = HEP-PH/0508265;%%

%\cite{Ball:2013bra}
\bibitem{Ball:2013bra}
  R.~D.~Ball, M.~Bonvini, S.~Forte, S.~Marzani, G.~Ridolfi and ,
  %``Higgs production in gluon fusion beyond NNLO,''
  arXiv:1303.3590 [hep-ph].
  %%CITATION = ARXIV:1303.3590;%%
  %1 citations counted in INSPIRE as of 01 Apr 2013

  %\cite{Pak:2011hs}
\bibitem{Pak:2011hs}
  A.~Pak, M.~Rogal and M.~Steinhauser,
  %``Production of scalar and pseudo-scalar Higgs bosons to next-to-next-to-leading order at hadron colliders,''
  JHEP {\bf 1109} (2011) 088
  [arXiv:1107.3391 [hep-ph]].
  %%CITATION = ARXIV:1107.3391;%%

%\cite{Anastasiou:2012kq}
\bibitem{Anastasiou:2012kq}
  C.~Anastasiou, S.~Buehler, C.~Duhr and F.~Herzog,
  %``NNLO phase space master integrals for two-to-one inclusive cross sections in dimensional regularization,''
  JHEP {\bf 1211} (2012) 062
  [arXiv:1208.3130 [hep-ph]].
  %%CITATION = ARXIV:1208.3130;%%

%\cite{Hoschele:2012xc}
\bibitem{Hoschele:2012xc}
  M.~Hoschele, J.~Hoff, A.~Pak, M.~Steinhauser and T.~Ueda,
  %``Higgs boson production at the LHC: NNLO partonic cross sections through order $\epsilon$ and convolutions with splitting functions to N$^3$LO,''
  arXiv:1211.6559 [hep-ph].
  %%CITATION = ARXIV:1211.6559;%%

%\cite{Anastasiou:2013srw}
\bibitem{Anastasiou:2013srw}
  C.~Anastasiou, C.~Duhr, F.~Dulat and B.~Mistlberger,
  %``Soft triple-real radiation for Higgs production at N3LO,''
  arXiv:1302.4379 [hep-ph].
  %%CITATION = ARXIV:1302.4379;%%

%\cite{Chetyrkin:1997un}
\bibitem{Chetyrkin:1997un}
  K.~G.~Chetyrkin, B.~A.~Kniehl and M.~Steinhauser,
  %``Decoupling relations to O (alpha-s**3) and their connection to low-energy theorems,''
  Nucl.\ Phys.\ B {\bf 510} (1998) 61
  [hep-ph/9708255].
  %%CITATION = HEP-PH/9708255;%%

%\cite{Schroder:2005hy}
\bibitem{Schroder:2005hy}
  Y.~Schroder and M.~Steinhauser,
  %``Four-loop decoupling relations for the strong coupling,''
  JHEP {\bf 0601} (2006) 051
  [hep-ph/0512058].
  %%CITATION = HEP-PH/0512058;%%

%\cite{Anastasiou:2010bt}
\bibitem{Anastasiou:2010bt}
  C.~Anastasiou, R.~Boughezal and E.~Furlan,
  %``The NNLO gluon fusion Higgs production cross-section with many heavy quarks,''
  JHEP {\bf 1006} (2010) 101
  [arXiv:1003.4677 [hep-ph]].
  %%CITATION = ARXIV:1003.4677;%%

%\cite{Furlan:2011uq}
\bibitem{Furlan:2011uq}
  E.~Furlan,
  %``Gluon-fusion Higgs production at NNLO for a non-standard Higgs sector,''
  JHEP {\bf 1110} (2011) 115
  [arXiv:1106.4024 [hep-ph]].
  %%CITATION = ARXIV:1106.4024;%%

%\cite{Pak:2010cu}
\bibitem{Pak:2010cu}
  A.~Pak, M.~Steinhauser and N.~Zerf,
  %``Towards Higgs boson production in gluon fusion to NNLO in the MSSM,''
  Eur.\ Phys.\ J.\ C {\bf 71} (2011) 1602
   [Erratum-ibid.\ C {\bf 72} (2012) 2182]
  [arXiv:1012.0639 [hep-ph]].
  %%CITATION = ARXIV:1012.0639;%%

  %\cite{Gonsalves:1983nq}
\bibitem{Gonsalves:1983nq}
  R.~J.~Gonsalves,
  %``Dimensionally Regularized Two Loop On-shell Quark Form-factor,''
  Phys.\ Rev.\ D {\bf 28} (1983) 1542.
  %%CITATION = PHRVA,D28,1542;%%
  
  %\cite{Kramer:1986sr}
\bibitem{Kramer:1986sr}
  G.~Kramer and B.~Lampe,
  %``Integrals For Two Loop Calculations In Massless Qcd,''
  J.\ Math.\ Phys.\  {\bf 28} (1987) 945.
  %%CITATION = JMAPA,28,945;%%
  
  %\cite{Gehrmann:2005pd}
\bibitem{Gehrmann:2005pd}
  T.~Gehrmann, T.~Huber and D.~Maitre,
  %``Two-loop quark and gluon form-factors in dimensional regularization,''
  Phys.\ Lett.\ B {\bf 622} (2005) 295
  [hep-ph/0507061].
  %%CITATION = HEP-PH/0507061;%%

%\cite{Vogt:2004mw}
\bibitem{Vogt:2004mw}
  A.~Vogt, S.~Moch and J.~A.~M.~Vermaseren,
  %``The Three-loop splitting functions in QCD: The Singlet case,''
  Nucl.\ Phys.\ B {\bf 691} (2004) 129
  [hep-ph/0404111].
  %%CITATION = HEP-PH/0404111;%%

%\cite{Moch:2004pa}
\bibitem{Moch:2004pa}
  S.~Moch, J.~A.~M.~Vermaseren and A.~Vogt,
  %``The Three loop splitting functions in QCD: The Nonsinglet case,''
  Nucl.\ Phys.\ B {\bf 688} (2004) 101
  [hep-ph/0403192].
  %%CITATION = HEP-PH/0403192;%%

\bibitem{Vogtwebsite}
  {http://www.liv.ac.uk/~avogt/hepres2.h}

%\cite{Remiddi:1999ew}
\bibitem{Remiddi:1999ew}
  E.~Remiddi and J.~A.~M.~Vermaseren,
  %``Harmonic polylogarithms,''
  Int.\ J.\ Mod.\ Phys.\ A {\bf 15} (2000) 725
  [hep-ph/9905237].
  %%CITATION = HEP-PH/9905237;%%
  
%\cite{Gehrmann:2001pz}
\bibitem{Gehrmann:2001pz}
  T.~Gehrmann and E.~Remiddi,
  %``Numerical evaluation of harmonic polylogarithms,''
  Comput.\ Phys.\ Commun.\  {\bf 141} (2001) 296
  [hep-ph/0107173].
  %%CITATION = HEP-PH/0107173;%%
  
  %\cite{Vollinga:2004sn}
\bibitem{Vollinga:2004sn}
  J.~Vollinga and S.~Weinzierl,
  %``Numerical evaluation of multiple polylogarithms,''
  Comput.\ Phys.\ Commun.\  {\bf 167} (2005) 177
  [hep-ph/0410259].
  %%CITATION = HEP-PH/0410259;%%
  
 %\cite{Maitre:2005uu}
\bibitem{Maitre:2005uu}
  D.~Maitre,
  %``HPL, a mathematica implementation of the harmonic polylogarithms,''
  Comput.\ Phys.\ Commun.\  {\bf 174} (2006) 222
  [hep-ph/0507152].
  %%CITATION = HEP-PH/0507152;%%
  
%\cite{Maitre:2007kp}
\bibitem{Maitre:2007kp}
  D.~Maitre,
  %``Extension of HPL to complex arguments,''
  Comput.\ Phys.\ Commun.\  {\bf 183} (2012) 846
  [hep-ph/0703052 [HEP-PH]].
  %%CITATION = HEP-PH/0703052;%%
 
  %\cite{Duhr:2011zq}
\bibitem{Duhr:2011zq}
  C.~Duhr, H.~Gangl and J.~R.~Rhodes,
  %``From polygons and symbols to polylogarithmic functions,''
  arXiv:1110.0458 [math-ph].
  %%CITATION = ARXIV:1110.0458;%%

%\cite{Buehler:2011ev}
\bibitem{Buehler:2011ev}
  S.~Buehler and C.~Duhr,
  %``CHAPLIN - Complex Harmonic Polylogarithms in Fortran,''
  arXiv:1106.5739 [hep-ph].
  %%CITATION = ARXIV:1106.5739;%%

 %\cite{Goncharov:1998}
 \bibitem{Goncharov:1998}
 A.~B.~Goncharov, 
 %``Multiple polylogarithms, cyclotomy and modular complexes,''
  Math. Research Letters, {\bf 5} (1998), 497--516 [arXiv:1105.2076].
    
 %\cite{Goncharov:2001}
  \bibitem{Goncharov:2001}
  A.~B.~Goncharov, 
  %``Multiple polylogarithms and mixed Tate motives,'' 
  (2001) [math/0103059v4].

%\cite{Duhr:2012fh}
\bibitem{Duhr:2012fh}
  C.~Duhr,
  %``Hopf algebras, coproducts and symbols: an application to Higgs boson amplitudes,''
  JHEP {\bf 1208} (2012) 043
  [arXiv:1203.0454 [hep-ph]].
  %%CITATION = ARXIV:1203.0454;%%

\bibitem{Goncharov-simple-Grassmannian} A.B.~Goncharov, 
  %``A simple construction of Grassmannian polylogarithms'', 
  [arXiv:0908.2238v3 [math.AG]].


\bibitem{Goncharov:2010jf}
  A.~B.~Goncharov, M.~Spradlin, C.~Vergu and A.~Volovich,
  %``Classical Polylogarithms for Amplitudes and Wilson Loops,''
  Phys.\ Rev.\ Lett.\  {\bf 105} (2010) 151605
  [arXiv:1006.5703 [hep-th]].
  %%CITATION = PRLTA,105,151605;%%

\bibitem{kitwebsite}
  http://www-ttp.particle.uni-karlsruhe.de/Progdata/ttp12/ttp12-45/

%\cite{Catani:2001ic}
\bibitem{Catani:2001ic}
  S.~Catani, D.~de Florian and M.~Grazzini,
  %``Higgs production in hadron collisions: Soft and virtual QCD corrections at NNLO,''
  JHEP {\bf 0105} (2001) 025
  [hep-ph/0102227].
  %%CITATION = HEP-PH/0102227;%%

%\cite{Olness:2009qd}
\bibitem{Olness:2009qd}
  F.~I.~Olness and D.~E.~Soper,
  %``Correlated theoretical uncertainties for the one-jet inclusive cross section,''
  Phys.\ Rev.\ D {\bf 81} (2010) 035018
  [arXiv:0907.5052 [hep-ph]].
  %%CITATION = ARXIV:0907.5052;%%
  %6 citations counted in INSPIRE as of 13 May 2013

  %\cite{Huber:2005yg}
\bibitem{Huber:2005yg} 
  T.~Huber and D.~Maitre,
  %``HypExp: A Mathematica package for expanding hypergeometric functions around integer-valued parameters,''
  Comput.\ Phys.\ Commun.\  {\bf 175}, 122 (2006)
  [hep-ph/0507094].
  %%CITATION = HEP-PH/0507094;%%

%\cite{Ridder:2012dg}
\bibitem{Ridder:2012dg}
  A.~G.~-D.~Ridder, T.~Gehrmann, E.~W.~N.~Glover and J.~Pires,
  %``Double Virtual corrections for gluon scattering at NNLO,''
  arXiv:1211.2710 [hep-ph].
  %%CITATION = ARXIV:1211.2710;%%

%%%%% below this line, references without actual citation in the text. %%%%%%

\end{thebibliography}
